\documentclass[referee]{aastex}
\usepackage{txfonts}
\usepackage{graphicx,subfigure}
\usepackage{epsfig}
\usepackage{epsf}

\usepackage{natbib}

\newcommand{\mic}{\,{\rm \mu m} } 
\newcommand{\XCO}{\rm X_{CO}} 

\begin{document}
\title{Statistical study of dust properties in LMC molecular clouds}
\author{D\'eborah Paradis \altaffilmark{1} \altaffiltext{1}{Spitzer Science Center, California Institute of Technology, Pasadena, CA 91125}, William T. Reach \altaffilmark{1}, Jean-Philippe Bernard \altaffilmark{2,3} \altaffiltext{2}{Universit\'e de Toulouse, UPS, CESR, 9 avenue du Colonel Roche,
F-31028 Toulouse, cedex 4, France} \altaffiltext{3}{CNRS, UMR 5187 F-31028, Toulouse, France}, Suzanne Madden \altaffilmark{4} \altaffiltext{4}{Service d'Astrophysique CEA, Saclay, 91191 Gif Sur Yvette Cedex, France}, Kazuhito Dobashi \altaffilmark{5} \altaffiltext{5}{Tokyo Gakugei University, Dept. of Astronomy and Earth Science, Koganei, Tokyo, 184-8501, Japan}, Margaret Meixner \altaffilmark{6} \altaffiltext{6}{Space Telescope Science Institute, 3700 San Martin Drive, Baltimore, MD 21218, USA},Toshikazu Onishi \altaffilmark{7} \altaffiltext{7}{Nagoya University, Dept. of Astrophysics, Chikusa-Ku, Nagoya, 464-01, Japan}, Akiko Kawamura \altaffilmark{7} and Yasuo Fukui \altaffilmark{7}}
\begin{abstract}

The objective of this paper is to construct a catalog providing the dust properties and the star formation efficiency (SFE) of the molecular clouds in the Large Magellanic Cloud (LMC). We use the infrared (IR) data obtained with the Spitzer telescope as part of the ``Surveying the Agents of a Galaxy's Evolution'' (SAGE) Legacy survey as well as the IRAS data. We also work with extinction (A$\rm _v$) maps of the LMC. A total of 272 molecular clouds have been detected in the LMC in a previous molecular survey, accounting for 230 giant molecular clouds and 42 smaller clouds. We perform correlations between the IR emission/extinction, and atomic and molecular gas tracers. We compare the atomic gas that surrounds the molecular cloud with the molecular gas in the cloud. Using a dust emission model, we derive the physical properties of dust in and outside the molecular clouds, such as the equilibrium temperature, composition, emissivity and extinction. We also determine the luminosity of the interstellar radiation field (ISRF) intercepted by the cloud, and the total IR luminosity from dust emission. The ratio of the IR luminosity to the gas mass traced by CO is used as an indicator of the SFE.  \\
Statistically, we do not find any significant difference in the dust properties between the atomic and the molecular phases. In particular we do not find evidence for a systematic decrease of the dust temperature in the molecular phase, with respect to the surrounding, presumably atomic gas. This is probably because giant molecular clouds are the sites of star formation, which heat the dust, while the smallest clouds are unresolved. The ratio between the infrared luminosity and the cloud mass ($\rm L^{Dust}_{TOT}/M_{gas}$) does not seem to correlate with $\rm M_{gas}$. The highest value of the ratio we derived is 18.1 $\rm L_{\odot}/M_{\odot}$ in the 30 Doradus region, which is known to be the most prominent star formation region of the LMC, while the most likely value is 0.5 and is representative of quiescent clouds. We provide a prescription to associate the various stages of star formation with its $\rm L^{Dust}_{TOT}/M_{gas}$.

\end{abstract}  


\section{Introduction}
Studying the dust properties in molecular clouds is important to understand the overall process of dust evolution between different phases, such as between the diffuse and the dense interstellar medium (ISM). Characterizing the dust properties of molecular clouds also sheds light on the star formation activity inside or at the periphery of the clouds. How the physical properties of molecular clouds evolve and how this process is related to the evolution of young stellar objects (YSOs) and other stellar cluster activity are not straightforward. Molecular clouds can be massive yet be quiescent in terms of star formation activity \citep{Lis01}. On the other hand, some clouds show evidence for significant internal heating or luminous internal sources resulting from high-mass star formation. The dust mass and abundance, the infrared (IR) luminosity and the cloud mass characterize the nature of the cloud, and allow us to quantify its star formation efficiency \citep{Mooney88, Scoville89, Sanders91, Deane94}. 

Comparing the dust associated with the atomic and molecular phases in clouds, we look for variations of the dust properties from the outside of the cloud to the inside. Indeed, we expect that dust evolves depending on the environment, as evidenced by chemical species depletion (essentially CO) \citep{Caselli99,Bacmann02}, a decrease of the IRAS $\rm I_{60}/I_{100}$ \citep{Laureijs91}, a decrease of the big grain (BG) equilibrium temperature, and an increase of the dust emissivity \citep{Stepnik03, Paradis09b}. These variations between diffuse and dense environments may be due to grain aggregation. The polycyclic aromatic hydrocarbons (PAHs) and very small grains (VSGs), with a size $>$0.4 nm and 1.2 nm, respectively,  dominate the emission in the near-infrared (NIR) whereas the BGs ($>$15 nm) are responsible for essentially all of the far-infrared (FIR) emission,  i.e. $\lambda \geq 100$ $\mic$ \citep{Desert90}. \citet{Stepnik03} suggested that the VSGs could stick on BG aggregates, inducing a decrease of the VSG relative abundance compared to BGs, and a decrease of the BG temperature due to more efficient emission by an aggregate than by isolated grains. These effects can be seen through changes in the dust temperature and therefore in the spectral energy distribution (SED).

The Large Magellanic Cloud (LMC) is an ideal laboratory to study molecular cloud properties due to its proximity \citep[$\simeq$ 50 kpc][]{Feast99,Leeuwen07} and  its almost face-on viewing angle. This allows studies of individual molecular clouds which do not suffer from crowding and source confusion. The ISM in the LMC is expected to be different from that in the Milky Way (MW), because of the lower metallicity \citep[1/2-1/3 that of the solar neighborhood;][]{Westerlund97}.
\citet{Cohen88} obtained the first CO map of the LMC using the southern CfA 1.2 m telescope at CTIO, at an angular resolution of 8.8$^{\prime}$ (130 pc at the distance of the LMC). Other observations at higher resolution were obtained from the ESO SEST Survey of the Magellanic Clouds, but only toward selected HII regions, dark clouds and IRAS infrared sources \citep[for instance][]{Israel86, Johansson94}. \citet{Fukui99} reported systematic observations of the overall molecular clouds of the LMC in the $^{12}$CO (J=1-0) transition, using the NANTEN, 4-m radio telescope, installed in Chile. A second survey of molecular clouds in the same transition was carried out with a higher sensitivity, covering an area of $\simeq$ 30 square degrees. This survey revealed the presence of a total of 272 molecular clouds, including 230 giant molecular clouds (MCs) and 42 smaller clouds \citep{Fukui08}. 164 of the 230 giant molecular clouds (GMCs) have radii ranging from 10 to 220 pc and virial masses ranging from $\rm 9 \times 10^3$ to $\rm 9 \times 10^6\,M_{\odot}$. 
Assuming virial equilibrium, \citet{Fukui08} estimated the CO to N$\rm _{H_2}$ conversion factor for each cloud in their survey, and found an average value of $\rm X_{CO}=(7 \pm 2) \times 10^{20}$ H/cm$^2$/(K km s$^{-1}$). \\

Recent studies have provided some evidence for variations in the gas and dust properties within molecular clouds in the LMC. \citet{Dobashi08} studied extinction and dust/gas ratio in the LMC molecular clouds, using correlations between extinction maps produced using star count in the NIR domain and atomic (HI) and molecular (CO) gas. They concentrated on 21 regions detected in both the A$\rm _v$ and the CO maps. They found an increase of the A$\rm _v$/N$\rm _H$ ratio from the outer regions of the LMC to the 30 Doradus star-forming region which they suggested could be due to an increase of the dust abundance close to star-forming regions or the presence of an additional gas component, not detected in HI nor in CO emission.
Molecular clouds in the LMC appear to be compact, and are likely to have a relatively extended CO-free H$_2$ envelope in agreement with the findings of \citet{Bernard08} based on the IR emission. Similar findings have been obtained for the Small Magellanic Cloud \citep[SMC,][]{Leroy09}. This possible additional molecular gas and how we take it into account in this study is discussed later in Section \ref{sec_add_gas}.
While \citet{Bernard08} did not find evidence for large variations in the BG equilibrium temperature towards molecular clouds compared to the surrounding atomic medium at 4$^{\prime}$ resolution, \citet{Paradis09a} found variations in the spatial distribution of the abundances of the different dust components across the LMC. They noted an increase of the PAH relative abundance in the old-population stellar bar, and around some molecular clouds. The VSGs also appear to be overabundant in a region around 30 Doradus and in a region near the center of the bar, and tend to follow star formation activity, whereas the PAH abundance essentially trace quiescent environments.
The objective of  this paper is to perform a statistical study of the dust properties in the atomic and molecular phase of each cloud of the LMC, in order to investigate the evolution of dust and determine how this is manifested in different environmental conditions.

Sections \ref{sec_obs} and \ref{sec_add_gas} summarize the data we use and the description of the FIR excess evidenced in previous studies.  In Section \ref{sec_corr} we present the method we applied to separate the IR emission arising from the molecular and atomic phases. Section \ref{sec_dust_prop} describes the dust emission model used to derive the dust properties for each molecular cloud, presented in Section \ref{sec_prop}. In Section \ref{sec_lum} and \ref{sec_sf} we focus on the luminosities (incident and emitted) computed for each cloud, and on the gas mass and star formation efficiency, respectively. Section \ref{sec_cl} is devoted to conclusions. 

\section{Observations} 
\label{sec_obs}
\subsection{IR data}
In order to trace the IR emission from dust, we use the Infrared Array Camera \citep[IRAC,][]{Fazio04} and the Multiband Imaging Photometer for Spitzer \citep[MIPS,][]{Rieke04} on board Spitzer as well as the IRIS \citep[Improved Reprocessing of the IRAS Survey, see][]{MamD05} data.
 IRAC observed at  3.6, 4.5, 5.8 and 8 $\mic$, with an angular resolution ranging from 1.6$^{\prime\prime}$ to 1.9$^{\prime \prime}$. MIPS provided images at 24, 70 and 160 $\mic$ at an angular resolution of 6$^{\prime \prime}$, 18$^{\prime
\prime}$ and 40$^{\prime \prime}$ respectively. The Spitzer data were obtained as part of the SAGE (``Surveying the Agents of a Galaxy's Evolution'') Spitzer legacy survey \citep{Meixner06}, covering the entire LMC ($\sim 7\degr \times7\degr$) at all the IRAC and MIPS bands.
We use the combined maps, at a pixel scale of 3.6$^{\prime \prime}$/pixel for IRAC, and 2.49, 4.8, and 15.6$^{\prime
\prime}$/pixel for the MIPS 24, 70, and 160 $\mic$ maps, respectively.

The IRAS satellite observed the entire LMC in four bands at 12, 25, 60 and 100 $\mic$ wavelength, with an angular resolution of 3.8, 3.8, 4.0 and 4.3$^{\prime}$, respectively. The advantages of the IRIS maps relative to the original IRAS maps reside in a better zodiacal light subtraction and calibration and zero level adjustments to match the DIRBE (Diffuse Infrared Background) data on large angular scales. 

\subsection{Extinction data} \label{sec_av}
The extinction map we use has been obtained using a star catalog from the SIRIUS camera on the InfraRed Survey Facility 1.4 m telescope \citep{Kato07}. The map has been constructed with an improved version of the NIR Color Excess method (NICE), initially introduced by \citet{Lada94}. A full description of these data and method is given by \citet{Dobashi09}. The NICE method measures the difference between the ``typical'' color of the stars located towards the cloud and toward a reference field. In the original and an early extension of the NICE method \citep[e.g.][]{Cambresy02}, this typical color value is taken to be the average or median star color value in each cell surrounding the cloud. Compared to the classical methods, the ``X-percentile'' method introduced by \citet{Dobashi09} does not assume the average or median color value but relies on the color of the X percentile reddest stars of each cell  (X=100$\%$ corresponds to the reddest star).  As shown by \citet{Dobashi09}, the method has an advantage to be robust against contamination by the foreground stars for high X values (e.g., X$>$50$\%$). In this paper, we applied their method to the SIRIUS catalog in the same way they did for the SMC, and derived color excess maps of E(J-H) and E(H-Ks) covering the entire LMC at various X$\%$. We set the angular resolution of the maps to be 2.6$^{\prime}$, the same as that of the CO map obtained using the NANTEN telescope (see Section \ref{sec_gas}). Among the resulting maps, we selected the color excess map of E(J-H) at X=80$\%$, because the map at this X value has a moderate noise level and reveals most of known clouds in the LMC. We then converted the color excess map into visual extinction A$\rm _v$ assuming the extinction law by \citet{Cardelli89} for R$\rm _v$=3.1:
\begin{equation}
A_v=10.9E(J-H) 
\end{equation}

\subsection{Gas tracers}
\label{sec_gas}
To trace the molecular phase and compare its distribution with that of the IR data, we use the second $^{12}$CO (J=1-0) survey obtained with the 4-m radio NANTEN telescope of Nagoya University at Las Campanas Observatory, Chile \citep[see][]{Fukui08}. The observations cover 6$\degr$ x 6$\degr$ toward the LMC, with a beam of 2.6$^{\prime}$ (about 40 pc at the distance of the LMC). The observations have been obtained over regions where molecular clouds have been detected with the first NANTEN survey. About 26900 positions have been observed. The observed grid spacing was 2$^{\prime}$ corresponding to around 30 pc. The measured rms noise is approximately 0.07 K at a velocity resolution of 0.065 km s$^{-1}$, with about 3 minutes of integration time per position. The 3-$\rm \sigma $ noise level on the integrated intensity is close to 1.2 K km s$^{-1}$. 
The individual clouds have been identified using the finding algorithm "fitstoprops" \citep{Rosolowsky06}. The clouds have been detected in the intensity data cube, which has been converted to a signal-to-noise-ratio \citep[see][for explanations on the cloud selection and description]{Fukui08}.
The CO to N$\rm _{H_2}$ conversion factor ($\rm X_{CO}$) is defined by: 
\begin{equation}
N_{H_2}=X_{CO}W_{CO}
\end{equation}
where $\rm W_{CO}$ is the integrated intensity of the CO emission.
The value of $\rm X_{CO}$ is not easily determined as it depends on the physical conditions of each environment like the metallicity, the ISRF strength and hardness which dissociates CO molecules, and it is directly linked to the efficiency of the H$_2$ formation on the dust grains. \citet{Fukui08} derived $\rm X_{CO}$ values under the assumption that clouds are in virial equilibrium ($\rm X_{CO}=M_{vir}/L_{CO}$, with $\rm M_{vir}$ the virial mass of the cloud and $\rm L_{CO}$ its CO luminosity). Note that the method is independent of the IR data and does not rely on any assumption about the dust properties. They found an average value of $\rm X_{CO}=(7 \pm 2) \times 10^{20}$ H/cm$^2$/(K km s$^{-1}$), with values for individual sources varying from cloud to cloud over a factor of 26, from 1.6$\times10^{20}$ to 4.2$\times10^{21}$ $\rm H_2/cm^2/(K\,km\,s^{-1})$. We use the $\rm X_{CO}$ values they determined for each cloud. In cases where the virial value could not be determined, we used the LMC average $\rm X_{CO}$ value. 

Neutral hydrogen is a major component of the gas in the LMC, with a total mass of $\rm 4\times 10^8\,M_\sun$ \citep{Bernard08} while the molecular mass traced by CO is only $\rm 5\times 10^7\,M_\sun$ \citep{Fukui08}. In order to study the distribution of the dust-to-gas (D/G) ratio, we use the HI map from \citet{Kim03}, which is a combination of interferometric data obtained with the Australia Telescope Compact Array (ATCA), at 1$^{\prime}$ (15 pc) angular resolution \citep{Kim03}, and the Parkes antenna, at 16.3$^{\prime}$ angular resolution \citep{Staveleysmith03}. The data cover a region of 10$\rm \degr$ $\times$ 12$\rm \degr$ in the LMC.
The integrated intensity map (W$\rm _{HI}$) was obtained by integrating the signal in the range 190 $\rm <v_{hel}<$386 km s$^{-1}$. 
Under the hypothesis that the gas is optically thin, we can derive the column density from the integrated intensity of the HI emission at 21 cm ($\rm W_{HI}$) using:
\begin{equation}
\label{eq_xhi}
N_{HI}=X_{HI}W_{HI}
\end{equation}
where the conversion factor is $\rm X_{HI}= 1.82\times10^{18}$ $\rm H_2/cm^2/(K\,km\,s^{-1})$ \citep{Spitzer78}. However the hypothesis of optically thin gas is questionable (see Section \ref{sec_add_gas}).

The IR and gas maps are taken directly from \citet{Bernard08} where the full description of the data processing can be found. Here we only give a brief summary of the treatment of the data. In their paper, point sources had been blanked on each IR map using archival catalogs (2MASS and SAGE catalogs). All maps, as well as the extinction map have been convolved to the IRIS 100 $\mic$ resolution (4$^{\prime}$) and have been projected on a common grid centered at $\rm \alpha_{2000}=05^h18^m48^s$ and $\rm \delta_{2000}=-68\degr42^{\prime}00^{\prime \prime}$, with a pixel size of $\rm 2^{\prime}$. The luminosity and mass determinations, have been performed at the NANTEN 2.6$^{\prime}$ resolution, to keep the best accuracy on the cloud size. 

\subsection{Foreground subtraction }
We have subtracted the Galactic foreground emission in the extinction map using:
\begin{equation}
A_v=A_v-\left ( \frac{A_v}{N_H} \right ) ^{gal} N_H^{gal} 
\end{equation}
where $\rm \left ( \frac{A_v}{N_H} \right ) ^{gal} $ is a reference relevant to the solar neighborhood and the plane of the Galaxy, taken to be 5.34$\times$10$^{-22}$ mag/(H cm$^{-2}$) \citep{Bohlin78} and $\rm N_H^{gal} $ is the Galactic HI column density map at 14$^{\prime}$ resolution, constructed by \citet{Staveleysmith03} by integrating the Parkes HI data in the velocity range of the MW emission (-64 and 100 km s$^{-1}$), away from the LMC velocities (see Section \ref{sec_gas}).
The foreground emission subtraction has also been applied to the IR maps, based on the same HI MW foreground map, as described in \citep{Bernard08}. We note that the resolution of the HI map used here for the foreground subtraction is coarser than that of our study carried at 4$^{\prime}$ resolution. Therefore Galactic residuals at scales between 4$^{\prime}$ and 14$^{\prime}$ are likely to remain in the maps. However this should not bias the results of our study which are based on a statistical study and should only contribute to increase the noise.

\section{Far-infrared excess}
\label{sec_add_gas}
\citet{Bernard08} and \citet{Dobashi08} detected the presence of a FIR and extinction excess in the inner regions of the LMC with respect to the dust/gas correlation measured in the outer regions. Two scenarios have been explored in \citet{Bernard08} to explain this excess: dust abundance variations or the presence of an additional gas component not traced by the available HI and CO measurements.
The first hypothesis would imply a lower dust abundance for all dust species in the molecular clouds, with respect to their surrounding atomic phase, which has never been systematically evidenced in the MW. Explaining such a systematic decrease of the abundance of all dust particles in molecular clouds would require more efficient dust production by evolved stars in peripheral regions of dense clouds. We consider this possibility unlikely. Color variations that can be attributed to BG sputtering have been observed at large scale in the LMC \citep{Bernard08}, which could affect as much as 13$\%$ of the BG mass in the 30 Doradus area, but which do not affect molecular cloud specifically. We therefore consider as a basic hypothesis that the BG abundance should remain statistically constant from the atomic to the molecular phase.
The second hypothesis invokes the existence of an additional gas phase. Such a phase has been evidenced in the MW through its gamma-ray contribution \citep{Grenier05}. In principle, this can result either from optically thick cold atomic gas or from dark molecular gas where H$_2$ is shielded from UV photodissociation while CO photodissociates \citep[e.g.][]{Wolfire10}.
In the LMC, this phase 
appears to be spatially correlated with the warm atomic phase \citep{Bernard08}. In the following we refer to this component as the FIR excess component. If not taken into account in our analysis, this FIR excess component would lead to systematically overestimating the dust abundance in the atomic phase surrounding the molecular clouds, by a factor close to 1.7.
We constructed a map of this component in the LMC, following the method proposed by \citet{Bernard08} (see their Equation 10). Unlike \citet{Bernard08}, however, we used the individual $\XCO$ values derived by \citet{Fukui08} for each molecular cloud when estimating its contribution to the total column density, and the LMC average $\XCO$ for those clouds not listed in \citet{Fukui08}.
Our map of the FIR excess, with a total mass of 3$\rm \times 10^{8}M_{\odot}$ is similar within 10$\%$ to that produced by \citet{Bernard08}. We note that the accuracy on the mass of the additional phase is uncertain and that this could affect the results of our study.
As the FIR excess component appears strongly correlated to HI and is still of unknown origin, we simply added this component to the HI map in the correlation study described in the following section.\\

\section{Correlations between IR/A$\rm _v$ and gas tracers}
\label{sec_corr}
We correlate the IR emission with gas tracers such as HI (which includes the FIR excess component as mentioned in section \ref{sec_add_gas}) and CO emission. The GMCs have radii ranging from 10 to 220 pc \citep[see][Table 3, for description of the GMCs properties]{Fukui08}. Compared to the definition of the molecular region provided in \citet{Fukui08}, we enlarged the cloud region by 4$^{\prime}$ in every direction, in order to include enough pixels with HI emission immediately surrounding the molecular cloud and little CO emission to use in the correlation study.
In order to avoid confusion from clouds on the same line of sight at different velocities, but located at the same pixel in the CO integrated map, we exclude pixels corresponding to significant CO emission by another cloud, from the correlation area. Correlations are performed only in regions with at least 15 valid pixels of 2$^{\prime}$ (equivalent radius larger than 4.4$^{\prime}$, assuming circular clouds). By this selection, we remove 13 clouds from the original catalog. In order to eliminate any possible residual zodiacal, Cosmic Infrared Background emission or small residual offsets in the maps, we subtract a background for each region in the IR, HI, CO and A$\rm _v$ maps. The background is computed as the median over a common background area, corresponding to the faintest half of the HI data. This step also ensures to have a null IR emission for a null column density.
We perform correlations using:
\begin{equation}
I_{\nu}(\lambda)=\frac{a_\nu(\lambda)}{X_{HI}}(N_H^{HI}+N_H^X)+\frac{b_\nu(\lambda)}{2X_{CO}}N_H^{CO}+c_{\nu}(\lambda)
\end{equation}
where $\rm I_{\nu}$ is the IR emission brightness at wavelength $\rm \lambda$, $\rm N_H^{HI}$, $\rm N_H^{X}$, and $\rm N_H^{CO}$ are the column density in the atomic, FIR excess, and molecular phases respectively. c$_{\nu}$ is a constant, which in principle should be close to zero since all maps used are background subtracted. It can also account for an additional gas phase, if not spatially correlated with the other gas components. It is also useful to assess the noise of the method. In the following, we  associate the FIR excess with the atomic component, and we refer to the sum of the atomic and FIR excess components as the atomic (or HI) phase. We however recall that the FIR excess component could well be molecular if associated with dark H$_2$ gas. We used the IDL (Interactive Data Language) linear regression function \texttt{regress} to search for $\rm a_\nu$ and $\rm b_\nu$ at each IR wavelength and for each molecular cloud.
The uncertainties on the data correspond to the standard deviation in a background region, divided by a mask which values range between 0 to 1 for pixels with or without contamination by the contribution from point sources, respectively. In this way we ensure that larger uncertainties are assigned to pixels contaminated by point source contribution. The results of the correlations for all molecular clouds, for the atomic ($\rm \frac{a_{\nu}(\lambda)}{X_{HI}}$) and the molecular ($\rm \frac{b_{\nu}(\lambda)}{2X_{CO}}$) phase, and the residual emission ($\rm c_{\nu}(\lambda)$) are given in Tables \ref{tab_hi}, \ref{tab_co} and \ref{tab_res} respectively. As expected, the histogram of the c($\lambda$) values is centered on zero, indicating no bias in the method. The scatter is about 5.2$\%$ of the total intensity.\\

In the same way as for the IR emission we assume that the total extinction in the visible is caused essentially by dust of the atomic and molecular phase. We perform the linear regression:
\begin{equation}
A_v=\frac{\alpha}{X_{HI}}(N_H^{HI}+N_H^X) + \frac{\beta}{2X_{CO}}N_H^{CO} + A_v^{0}
\end{equation}
where $\rm \alpha$ and $\beta$ are the free parameters and $\rm A_v^0$ is the residual extinction. Uncertainties on the A$\rm _v$ data correspond to the noise level on the Av map, described in \citet{Dobashi09}. The results of the correlations associated with the atomic ($\rm \frac{\alpha}{X_{HI}}=\left ( \frac{A_v}{N_H} \right )^{HI}$) and molecular ($\rm \frac{\beta}{2X_{CO}}=\left ( \frac{A_v}{N_H} \right )^{CO}$) phases, and the residual extinction ($\rm A_v^0$) are given in Tables \ref{tab_results_hi} and \ref{tab_results_co}. The $\rm A_v^0$ distribution does not evidence any bias in the method as its histogram is centered on 0, and its scatter is lower than 2.1$\%$ of the total extinction.

\section{Dust emission modeling}
\label{sec_dust_prop}
The SEDs of the correlation coefficients $\rm a_\nu$ and $\rm b_\nu$ are equivalent to the emission spectrum of dust associated with the atomic and molecular phases, for a given column density. We then apply the photometry correction to the correlation coefficients at the IRAC wavelengths, by multiplying them by 0.737, 0.772, 0.937 and 0.944 (from 3.6 to 8 $\mic$), to account for the difference between calibration on point-source and extended sources \citep{Reach05}. We fit the spectra of each phase with an updated version of the \citet{Desert90} model as implemented in the Dustem code \citep{Compiegne10}. The dust composition used here is that proposed by \citet{Desert90} and takes into account three dust grain populations: PAHs, VSGs mainly composed of carbonaceous material, and BGs composed of silicate.
We allow several parameters to vary: the ISRF intensity (X$\rm _{ISRF}$), assuming the same ISRF spectral shape as in the solar neighborhood, the mass abundances relative to hydrogen of each dust specie, noted as $\rm Y_{BG}$, $\rm Y_{VSG}$, and $\rm Y_{PAH}$, for the BG, VSG, and PAH, respectively. We note that, if the ISRF is actually harder than the solar neighborhood spectrum assumed here, the relative abundances of the small particles (PAHs and VSGs) would be decreased. On the contrary, the BG abundance would not differ much, since BG absorb their energy over the entire range of the radiation field spectrum. A significant 70 $\mic$ emission excess with respect to the model predictions has been evidenced in some regions of the LMC by \citet{Bernard08}. This should not be confused with the FIR excess, since they have different spatial distribution in the LMC and may be due to completely different processes. They suggested that it could be explained by a flattening in the VSG size distribution. \citet{Galliano03, Galliano05} also allowed the dust size distribution to vary in the SED models of a sample of low metallicity dwarf galaxies to fit this excess around 70 $\mic$ and did indeed require a different value compared to our Galaxy. In regions with strong 70 $\mic$ excess, such as within 1$\degr$ from 30 Doradus, the excess can impact the determination of X$\rm _{ISRF}$ and consequently of the dust temperature. However, since the origin of the excess is still not fully understood, we decided not to account for it by changing the VSG size distribution and kept the standard size distribution proposed by \citet{Desert90}. Instead, in order to ensure an accurate determination in the ISRF intensity, we force the model to reproduce the observations at 100 and 160 $\mic$ by increasing the weight of these data points in the SED modeling. We note that this may result in overestimating X$\rm _{ISRF}$ and the dust temperature and correspondingly underestimating the dust abundances in the 70 $\mic$ excess regions. However, since this excess is correlated mostly with HI \citep[see][]{Bernard08}, thus contributing to the atomic phase, this should not affect the results of the correlation for the molecular phase. Results of the modeling are given in Tables \ref{tab_results_hi} and  \ref{tab_results_co} for the atomic and molecular phases, respectively. Null values of the results are not given in the tables.

\section{Dust properties}
\label{sec_prop}
\subsection{Dust Temperature}
\label{sec_temperature}
Assuming that dust emission follows a greybody emission ($\rm I_{\nu} \propto \epsilon_0 \left (\frac{\lambda}{\lambda_0} \right )^{-\beta} B_{\nu}(\lambda,T_d)$, with $\epsilon_0$ the emissivity at wavelength $\lambda_0$, $\rm B_{\nu}$ the Planck function, $\rm T_d$ the dust temperature and $\beta$ the emissivity spectral index), the ISRF intensity is related to the BG equilibrium temperature through:
\begin{equation}
\label{eq_isrf}
\frac{X_{ISRF}^{LMC}}{X_{ISRF}^{\odot}}=\left ( \frac{T_d^{LMC}}{T_d^{\odot}} \right )^{4+\beta}
\end{equation}
where X$\rm _{ISRF}^{\odot}$ is the ISRF scaling factor in the solar neighborhood (X$\rm _{ISRF}^{\odot}\simeq 1$) and the dust temperature in the solar neighborhood ($\rm T_d^{\odot}$) is about 17.5 K \citep{Boulanger96} assuming $\beta=2$. $\rm X_{ISRF}^{LMC}$ is a free parameter of the model used and is deduced from the modeling of the SED for individual clouds. Dust temperatures are then computed using equation \ref{eq_isrf}. Note that the derived $\rm T_d$ is an average temperature, in the sense that the model in fact incorporates a distribution of temperatures, due to the existence of a dust size distribution. The spectral index is taken equal to $\beta=2$ in the dust model we use. We disregarded clouds showing negative $\rm a_\nu$ or $\rm b_\nu$ values at 100 or 160 $\mic$ for at least one of the gas phases, which induces a wrong determination of the ISRF intensity and therefore of the dust temperature. This selection removed 70 clouds from the original catalog, leaving 189 clouds. Temperature values, both associated with dust in the atomic and in the molecular phases are given in Tables \ref{tab_results_hi} and \ref{tab_results_co}, respectively. 

\begin{figure*}
\begin{center}
\includegraphics[width=8cm]{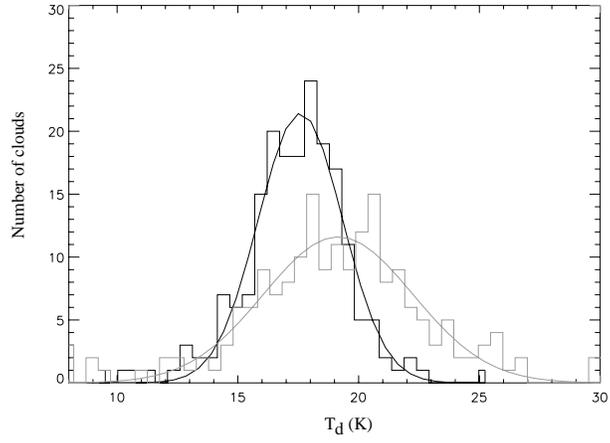}
\caption{Temperature histograms in the atomic (black) and molecular phase (grey).\label{fig_temp}}
\end{center}
\end{figure*}

Figure \ref{fig_temp} shows the histograms of the dust temperature for each component of the gas.
The most likely value of the temperature in the atomic and molecular phases, deduced from the Gaussian fit of the histograms are respectively $\rm T^{HI}_{d}$=17.6$\pm$1.8 K and $\rm T^{CO}_{d}$=19.2$\pm$3.1 K, with uncertainties corresponding to the 1-$\sigma$ width of the histograms.
Therefore, on average, dust in the molecular phase appears slightly warmer (by 1.6 K) than dust associated with the atomic phase, although the difference is not statistically significant. A similar increase of the dust temperature inside dense clouds compared to the diffuse ISM has already been observed by \citet{Stanimirovic00} in the SMC. A comparison between dense clouds in the LMC and the SMC can be found in \citet{vanloon10}.
A total of 47 clouds (25$\pm 4\%$) have significantly colder dust in the molecular phase than in the atomic phase, taking into account the uncertainties on the temperature derived from uncertainties on the $\rm X_{ISRF}$ determination ($T_d^{CO} + \Delta T_d^{CO} < T_d^{HI} - \Delta T_d^{HI}$). However, 104 clouds (55$\pm 6\%$) show significantly warmer dust in the molecular phase than in the surrounding atomic phase, considering the uncertainties on the temperature. The rest of clouds (20$\pm 3\%$) have the same temperature in both phases, within the error bars. A two-sided Kolmogorov-Smirnov (KS) test indicates that the probability of the temperature distributions in the two phases being drawn from the same parent population is 2.52$\times\,10^{-12}$. Therefore, the two distributions are significantly different.
Figure \ref{fig_temp} also highlights the fact that the histogram shapes for each phase are different: following the Gaussian fit, the full width at half maximum is larger for the molecular phase than for the atomic phase: 7.3 K and 4.1 K, respectively. This reflects the fact that the uncertainties on the dust temperature are most of the time higher in the molecular phase. Indeed, the correlation coefficients are determined with better accuracy in the atomic phase according to the uncertainties on the coefficients. 

From radiative transfer arguments alone, we would expect dust in the dense molecular phase to be colder than in the surrounding atomic medium. In the LMC this is the case for only 25$\%$ of the clouds.
However, active star formation inside the molecular region can in principle lead to a reverse situation. Note also that star formation at the periphery of the molecular cloud would also lead to such a situation, due to the limited angular resolution of this study (4$^{\prime}$ or 60 pc at the distance of the LMC).
The fact that in 75$\%$ of the cases the observed temperatures are comparable or warmer within uncertainties in the molecular phase compared to the atomic one, suggests that star formation actually occurs in or near the molecular clouds within our resolution, for a large number of molecular clouds.

\begin{figure*}
\begin{center}
\includegraphics[width=14cm]{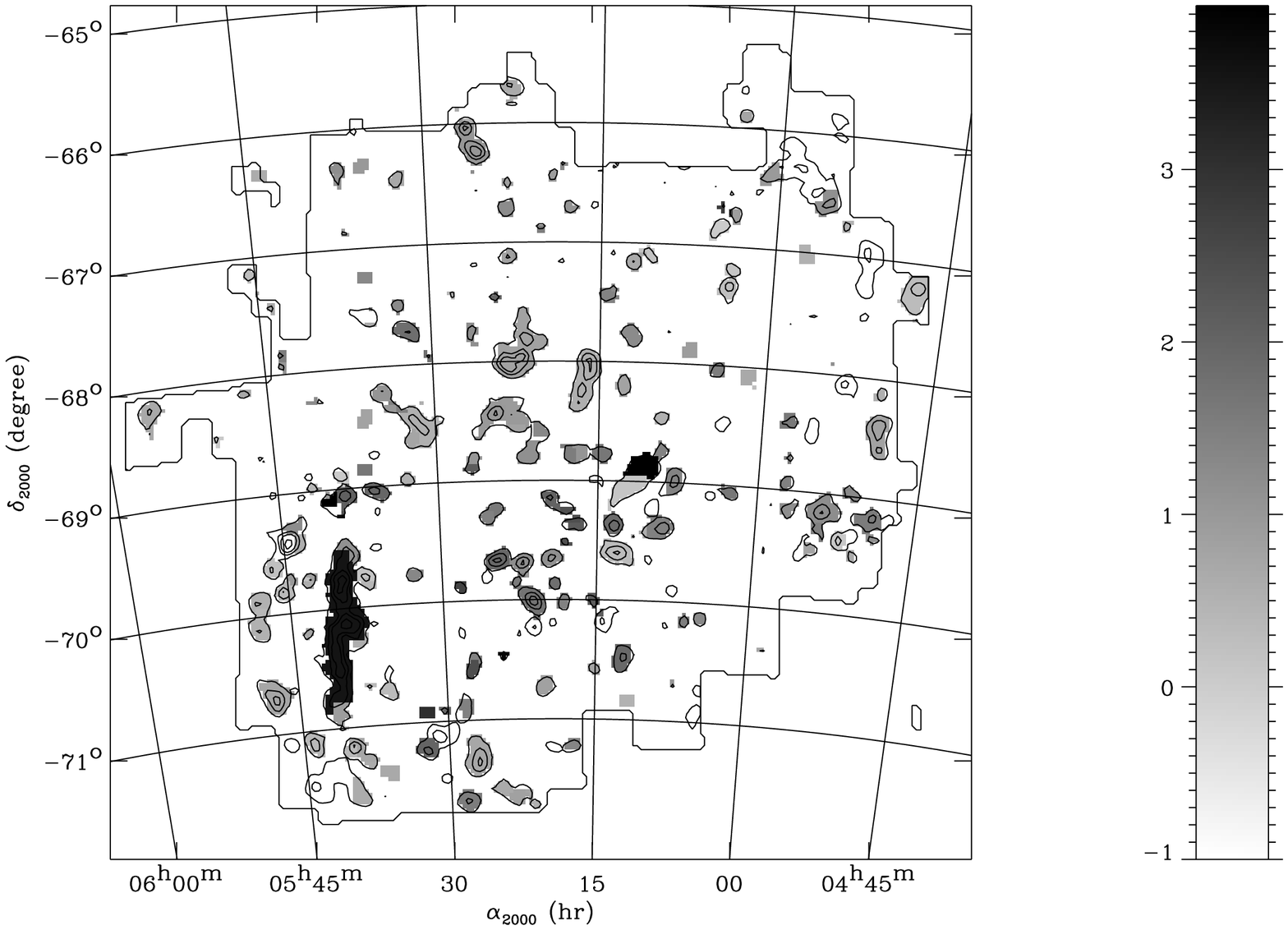}
\caption{Map of the intensity of the ISRF in the atomic phase, overlaid with the NANTEN $^{12}$CO (J=1-0) integrated intensity contours convolved to the 4$^{\prime}$ resolution, at 0.5, 2, 4 and 8 K km s$^{-1}$. \label{fig_xisrf_hi}}
\end{center}
\end{figure*}

\begin{figure*}
\begin{center}
\includegraphics[width=14cm]{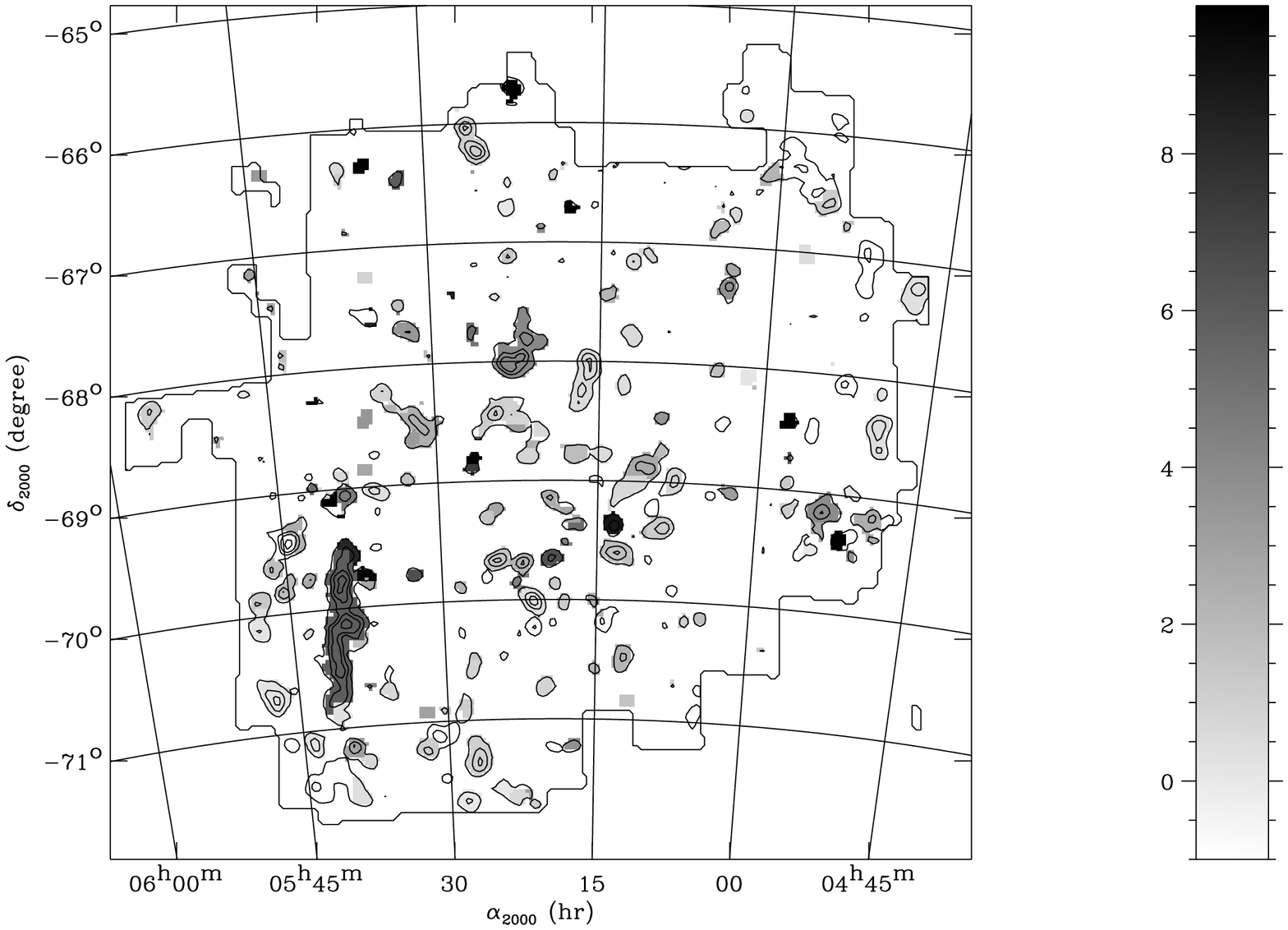}
\caption{Map of the intensity of the ISRF in the molecular phase, overlaid with the NANTEN $^{12}$CO (J=1-0) integrated intensity contours convolved to the 4$^{\prime}$ resolution, at 0.5, 2, 4 and 8 K km s$^{-1}$. \label{fig_xisrf}}
\end{center}
\end{figure*}

Figures \ref{fig_xisrf_hi} and \ref{fig_xisrf} show the spatial distribution of X$\rm _{ISRF}$ in each phase. Although the dust temperatures in both phases of each molecular cloud are quite similar statistically, it can be seen that large variations of the radiation field intensity exist from cloud to cloud. In particular, large X$\rm _{ISRF}$ in both phases are found near the 30 Doradus region and along the main molecular ridge south of 30 Doradus. Other large clouds located away from the main star-forming region, in the outer regions of the LMC exhibit lower values of X$\rm _{ISRF}$ in both phases. This indicates that most of the observed variations are due to the distribution of heating sources with respect to the clouds, and not to the attenuation by radiative transfer in the clouds. When smaller clouds are considered, there is no clear correlation between X$\rm _{ISRF}$ in the atomic and in the molecular phase. 

\subsection{Dust Abundances}
\label{sec_ab}

Absolute dust abundances in the LMC molecular clouds and their surrounding are significantly lower than those of the Galaxy, for all dust components. It is interesting to compute the total D/G ratio, to be compared with that in our Galaxy. The dust abundances defined in Section \ref{sec_dust_prop} correspond to the dust mass relative to hydrogen. Therefore D/G is given by:
\begin{equation}
D/G=Y_{PAH}+Y_{VSG}+Y_{BG}
\end{equation} 
Values of the D/G ratio are presented in Tables \ref{tab_results_hi} and \ref{tab_results_co}, for each phase.
The distribution of D/G in the atomic and molecular phase has a main peak at $(2.7^{+1.1}_{-1.6} )\times10^{-3}$ and $(2.2 ^{+1.5}_{-0.9})\times 10^{-3}$ respectively. The corresponding value in the solar neighborhood is $5.8 \times 10^{-3}$ \citep{Bernard08}.
Therefore the dust to gas mass ratio is found to be around 1/2.2, which is in fair agreement with the lower metallicity of the LMC compared to the solar neighborhood ($\rm Z_{LMC}=1/2-1/3Z_{\odot}$). 

\begin{figure*}
\begin{center}
\includegraphics[width=8cm]{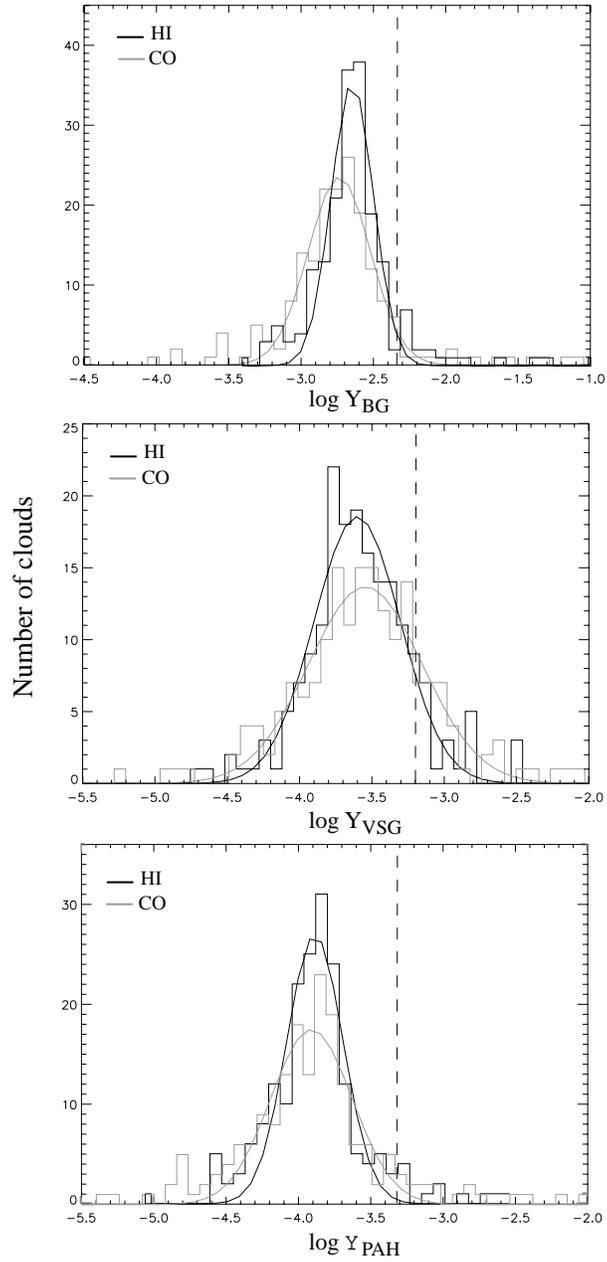}
\caption{Histograms of the dust abundances in the atomic phase in black and molecular phase in grey. From top to bottom: BG, VSG and PAH abundances. 
The dashed vertical lines are the abundances derived in \citet{Bernard08} for the solar neighborhood.\label{fig_hist_y}}
\end{center}
\end{figure*}

Figure \ref{fig_hist_y} shows the histograms of the abundances for each dust component for the atomic and molecular phase. We also show the reference values from the solar neighborhood determined by \citet{Bernard08} ($\rm Y_{BG}=4.68\times 10^{-3}$, $\rm Y_{VSG}=6.38\times 10^{-4}$ and $\rm Y_{PAH}=4.83\times 10^{-4}$) who used the same dust model and fitting approach.
The derived dust abundances are similar in the molecular and in the atomic phase. The two BG abundance histograms are expected to peak at similar values due to the hypotheses we made when constructing the FIR excess map. Observations of some quiescent and cold molecular clouds in the solar neighborhood indicate both low temperatures and high emissivities (or higher BG abundances), which are both evidences for dust aggregation \citep{Stepnik03, Paradis09b}. According to aggregation models, the temperature decrease can be caused by emissivities being larger for aggregates than for individual dust grains by a factor of $\sim$ 3. We note that although aggregation has been observed in several nearby MW molecular clouds \citep[e.g.][]{Bernard99, Stepnik03}, it is by no means a systematic effect. In the MW, clouds showing signs of aggregation also appear to be extremely quiescent with very narrow molecular lines \citep[see for instance][]{Pagani09}. If grain aggregation occurs in molecular clouds of the LMC, we would expect to detect it as a lower dust temperature. We would also expect to see higher BG abundances toward the molecular clouds, if the dust content is not significantly affected by other processes such as star formation in the cloud or its vicinity (see discussion in Section \ref{sec_add_gas}). We cannot conclude concerning the BG aggregation in molecular clouds of the LMC by only comparing the BG abundances in each phase, since the FIR excess map has been constructed by reconciling the emissivities of each phase at 160 $\mic$. However the fact that we do not see a systematic decrease of the dust temperature in the molecular phase could indicate that aggregation is not systematically taking place in the LMC. Our result may indicate that most LMC molecular clouds are actively forming stars and lack the low turbulence conditions necessary for dust aggregation. We note that it could also result from the limited angular resolution of our analysis, which amplifies the confusion between the emission associated with the molecular clouds and that of the surrounding diffuse medium. 
Inspection of Figure \ref{fig_hist_y} also shows that the abundance histogram is broader for the VSG component than for the BG and PAH components in both gas phases. The reason for this behavior is unclear, but it may indicate that the VSGs are actually subjected to more efficient processing in the ISM than the smallest (PAH) and largest (BG) dust particles. This may result from shattering processes from larger grains, which have been invoked in the LMC to explain the 70 $\mic$ excess.

\begin{figure*}
\begin{center}
\includegraphics[width=8cm]{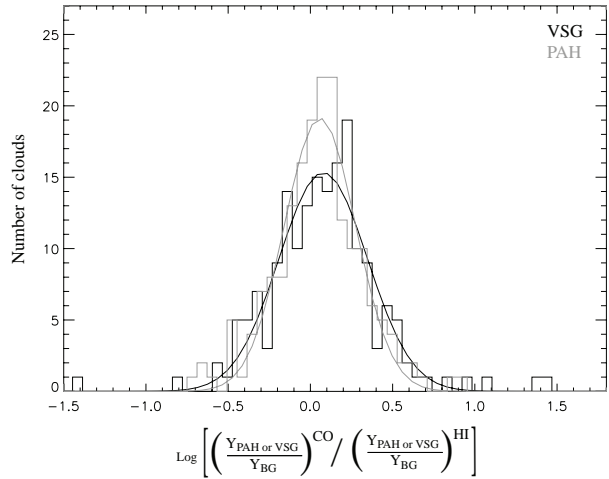}
\caption{Histograms of the logarithm of the $\rm \frac{Y_{PAH}}{Y_{BG}}$ (grey) and $\rm \frac{Y_{VSG}}{Y_{BG}}$ (black) ratios between the molecular and the atomic phase.  \label{fig_vsg_pah}}
\end{center}
\end{figure*}

\begin{figure*}
\begin{center}
\includegraphics[width=8cm]{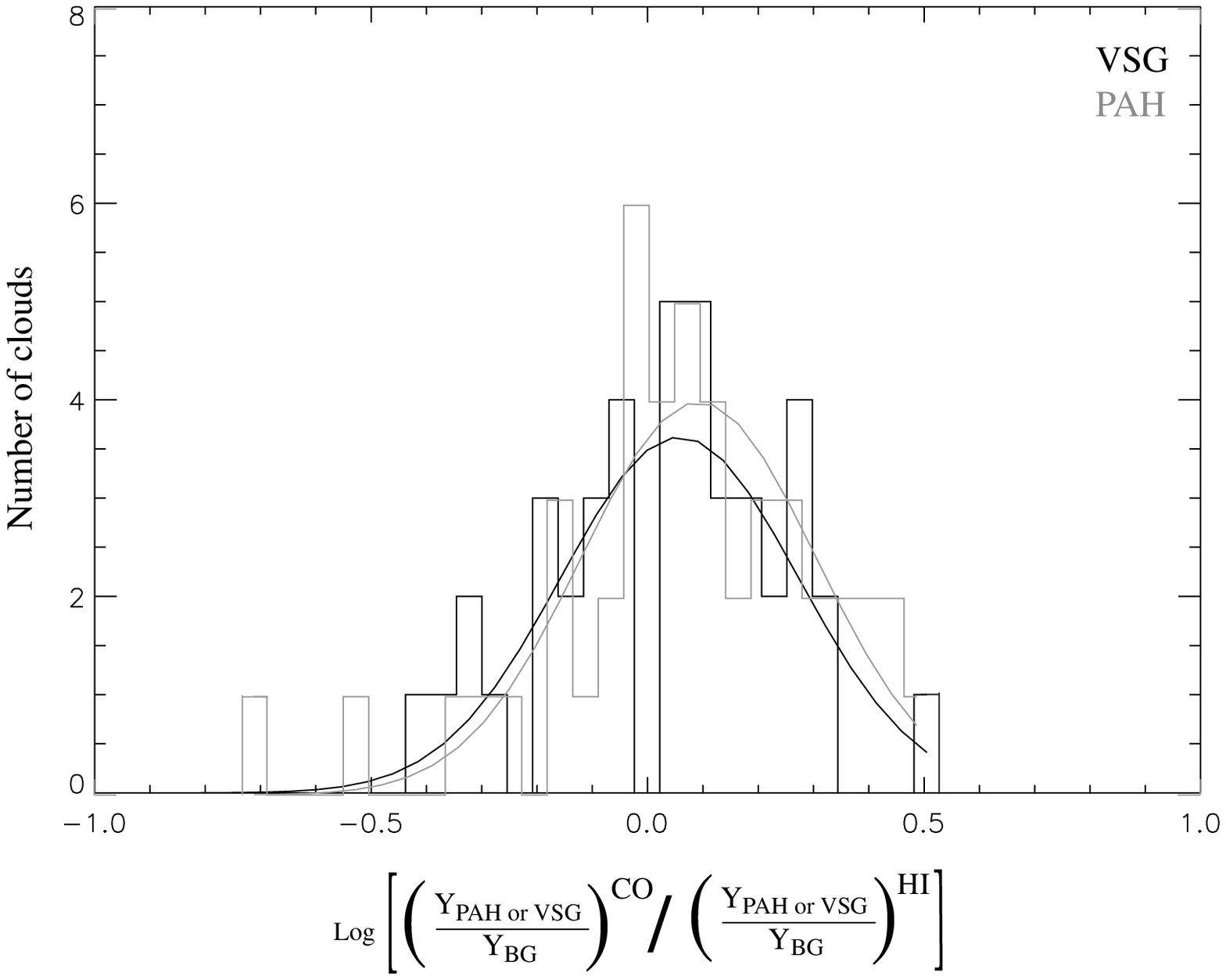}
\caption{Histograms of the logarithm of the $\rm \frac{Y_{PAH}}{Y_{BG}}$ (grey) and $\rm \frac{Y_{VSG}}{Y_{BG}}$ (black) ratios between the molecular and the atomic phase, for clouds which dust temperature in the molecular phase is colder than in the atomic phase.\label{fig_vsg_pah_thi_gt_tco}}
\end{center}
\end{figure*}

\begin{figure*}
\begin{center}
\includegraphics[width=8cm]{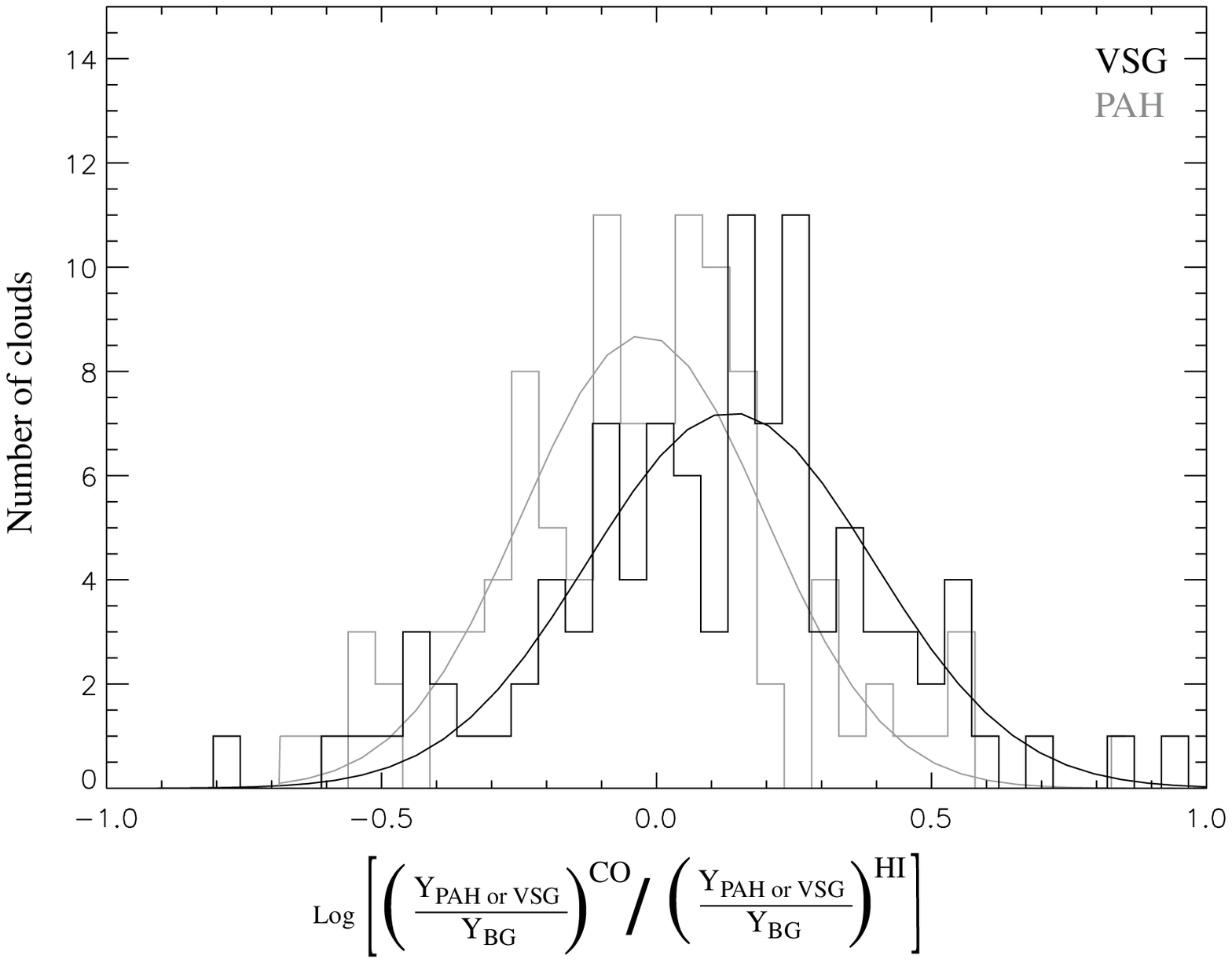}
\caption{Histograms of the logarithm of the $\rm \frac{Y_{PAH}}{Y_{BG}}$ (grey) and $\rm \frac{Y_{VSG}}{Y_{BG}}$ (black) ratios between the molecular and the atomic phase, for clouds which dust temperature in the molecular phase is warmer than in the atomic phase.\label{fig_vsg_pah_tco_gt_thi}}
\end{center}
\end{figure*}

If some evolution in the VSG and/or PAH properties occurs in molecular clouds, we would expect to see some variations in the VSG and/or PAH relative abundance compared to BGs, from the atomic to the molecular phase of the clouds. VSG and/or PAH aggregation would induce a decrease of the relative abundance in the molecular phase. To determine any possible changes in the dust abundances between the two phases, we plot in Figure \ref{fig_vsg_pah} the histogram of the VSG and PAH relative abundance ratios between the molecular and the atomic phase for all clouds. Values of this ratio are given in Tables \ref{tab_results_hi} and \ref{tab_results_co}, for the atomic and molecular phase, respectively. Statistically we can see that the peaks of the histograms of the logarithm of the ratios for the PAH and VSG components are very close, with a value around 0.07 (corresponding to a ratio of 1.17$^{+0.74}_{-0.45}$), indicating no significant change in the PAH and VSG relative abundance between the two phases, taking into account the large dispersion of the histograms. 
We conclude that statistically, in the majority of the clouds, there is no apparent evolution in the PAH and VSG properties between the atomic and the molecular phase at the 4$^{\prime}$ resolution of our analysis.
Figures \ref{fig_vsg_pah_thi_gt_tco} and \ref{fig_vsg_pah_tco_gt_thi} present the same histograms in the cases where the dust temperature in the molecular phase is significantly colder than that in the atomic phase, and in the opposite case. Figure \ref{fig_vsg_pah_thi_gt_tco} shows that in the cold molecular phase, with a median temperature of 16.0 K, there appears to be a 9$\%$ increase of the PAH relative abundance compared to that of the VSGs. But taking into account the dispersion of the abundances (full width at half maximum equal to 0.49 in logarithm), the increase of the PAH relative abundance is not significant. Figure \ref{fig_vsg_pah_thi_gt_tco} also shows that there is no change in the VSG relative abundance either in the cold molecular phase. A two-sided KS test on both the PAH and the VSG distributions, resulted in a probability of 0.302, which confirms that the two histograms are not significantly different.

In Figure \ref{fig_vsg_pah_tco_gt_thi}, the warm molecular phase, with a median temperature of 21.0 K, shows a statistically significant increase of the VSG relative abundance, close to 40$\%$ and no change of the PAH relative abundance. A two-sided KS test gives a probability of 3.2$\times$10$^{-4}$, indicating that the two distributions are significantly different.
The apparent increase of the VSG abundance could reveal in-situ production of VSGs associated with star formation inside or near the cloud. The resulting interaction between the stellar winds or outflows and the surrounding medium could produce VSG through erosion of BG. This process would occur inside or at the surface of the molecular clouds, and at the 4$^{\prime}$ resolution of our study, would appear correlated to the molecular phase.

\subsection{Dust optical depth}
\label{sec_dep}
Knowing the BG equilibrium temperature of dust in each phase of the molecular clouds, deduced from Section \ref{sec_temperature}, we computed the dust emissivity at 160 $\mic$ associated with the atomic ($\rm \epsilon_{\nu}^{HI}(\lambda)$) and molecular ($\rm \epsilon_{\nu}^{CO}(\lambda)$) phases respectively, for each cloud using: 
\begin{equation}
\epsilon_{\nu}^{HI}(\lambda)=\left ( \frac{\tau}{N_H} \right )^{HI}=\frac{a_{\nu}(\lambda)}{X_{HI}B_{\nu}(T_d^{HI})}
\end{equation}
and
\begin{equation}
\epsilon_{\nu}^{CO}(\lambda)=\left ( \frac{\tau}{N_H} \right )^{CO}=\frac{b_{\nu}(\lambda)}{2X_{CO}B_{\nu}(T_d^{CO})}
\end{equation}
where $\tau$ is the dust optical depth. In the same way we define the $\rm A_v/N_H$ ratio in the atomic and molecular phases:
\begin{equation}
\left ( \frac{A_v}{N_H} \right )^{HI} =\frac{\alpha}{X_{HI}}
\end{equation}
and 
\begin{equation}
\left (\frac{A_v}{N_H} \right )^{CO}=\frac{\beta}{2X_{CO}}.
\end{equation}
Values of the dust emissivity (or $\rm \tau/N_H$) and $\rm A_v/N_H$ associated with the atomic and molecular phase are given in Tables \ref{tab_results_hi} and \ref{tab_results_co}. Negative values of the $\rm A_v/N_H$ are not given in the tables.\\

In the solar neighborhood, \citet{Boulanger96} measured an emissivity value in the diffuse medium of $10^{-25}$ cm$^2$/H at 250 $\mic$ assuming a spectral index $\beta$ equal to 2 which was consistent with their data. We can thus infer the dust emissivity at wavelength $\lambda$ using 
\begin{equation}
\frac{\tau}{N_H}(\lambda)=1 \times 10^{-25} \left ( \frac{\lambda}{250\,\mu m} \right ) ^{-2}.
\end{equation} 
Therefore the derived solar neighborhood dust emissivity at 160 $\mic$ is 2.44$\times$10$^{-25}$ cm$^2$/H.
A reference for the $\rm A_v/N_H$ ratio in the solar neighborhood and in the plane of the Galaxy is 5.34$\times$10$^{-22}$ mag/(H cm$^{-2}$) \citep{Bohlin78}.
The histograms of the logarithm of $\rm \tau_{160}/N_H$ and $\rm A_v/N_H$ are presented in Figure \ref{fig_all_hist}, panels A and B respectively. We note that the peak position of the emissivity histograms for the atomic and molecular phases are in relatively good agreement. The peak position of the $\rm A_v/N_H$ histogram in the molecular phase is higher than in the atomic phase. However, due to the large dispersion in the $\rm (A_v/N_H)^{CO}$ values, the discrepancy between the phases is not significant. The most likely values for the emissivity and the $\rm A_v/N_H$ ratio for the atomic phase, deduced from the Gaussian fits are (1.0$^{+0.3}_{-0.3}$)$\times$10$^{-25}$ cm$^2$/H and (6.3$^{+4.5}_{-2.6}$)$\times$ 10$^{-23}$ mag/(H cm$^{-2}$) respectively.
The median emissivity value for the LMC is therefore 2.4 times lower than the solar neighborhood value. \citet{Bernard08} computed the emissivity in the external regions of the LMC and found  $\rm \tau_{160}/N_H\simeq8.8\times$10$^{-26}$ cm$^2$/H. Our median value is in agreement with their result, given the uncertainty on both values. However, the median $\rm A_v/N_H$ in the atomic phase appears to be 8.5 times lower than the solar reference.  The origin of the discrepancy, a factor 3.5 on the peak values between $\rm A_v/N_H$ and $\rm \tau_{160}/N_H$ is unclear. The overall $\rm \tau_{160}/A_v$ in the LMC is consistent with the MW value. So the difference observed here is only seen at small scales around clouds. It could be due to underestimating extinction in constructing $\rm A_v$ maps from star counts at small scale around clouds. This could arise either from improperly taking into account the effect of the gas and star mixing, or from noise at small scale in $\rm A_v$ maps.

The most likely values of the dust properties described in Section \ref{sec_prop} are summarized in Table \ref{tab_avg}.

\section{Luminosities }
\label{sec_lum}
We compute the luminosity intercepted by each cloud from the diffuse ISRF, L$\rm _{ISRF}$, assuming that all heating photons are absorbed in the cloud. This excludes the extra heating that may be caused by star formation in the cloud. We then compute the total IR luminosity emitted by dust L$\rm _{TOT}^{Dust}$ in the cloud.
The ratio of these two quantities is then used as an indication that clouds are either translucents (L$\rm _{TOT}^{Dust}<L_{ISRF}$) or optically thick (L$\rm _{TOT}^{Dust} \simeq L_{ISRF}$) and already significantly heated by star formed in or near the cloud (L$\rm _{TOT}^{Dust}>L_{ISRF}$).

Following \citet{Keene80} and assuming the cloud to be a sphere, the absorbed ISRF luminosity can be written:
\begin{equation}
L_{ISRF}= 4 \pi ^2R^2I_{ISRF}= 4 \pi N_{pix} D^2 \Omega_{pix} I_{ISRF}
\end{equation} 
where R is the cloud dimension, $\rm N_{pix}$ is the number of pixels, D is the distance of the cloud, $\rm \Omega_{pix}$ is the pixel solid angle and $\rm I_{ISRF}$ is the ISRF intensity in ergs cm$\rm ^{-2}$ s$\rm ^{-1}$, defined as:
\begin{equation}
4\pi I_{ISRF}= Uc
\end{equation}
with c the speed of light and U the energy density. We assumed U=0.5 eV $\rm cm^{-3}$ \citep{Allen00}, which is a representative value for diffuse HI clouds of our Galaxy. This Galactic value can be reasonably applied to the LMC since dust temperatures in the HI phase of both galaxies appear similar. We define the total luminosity emitted by dust as:
\begin{equation}
L_{TOT}^{Dust}= L_{BG}+L_{VSG}+L_{PAH}
\end{equation}
where $\rm L_{BG}$, $\rm L_{VSG}$ and $\rm L_{PAH}$ are respectively the luminosities from the BG, VSG and PAH dust components, computed as the integral of the model over the wavelength range 8 - 160 $\mic$. We note that this definition slightly underestimates the luminosity, compared to the bolometric luminosity $\rm L_{bol}$. For the clouds in our sample, the relation is approximately $L_{bol}=1.13\,L_{TOT}^{Dust}+2.2\,10^4\,L_{\odot}$. The luminosity for each dust component j is defined as:
\begin{equation}
\label{eq_lj}
L_j=4 \pi D^2 \Omega_{pix} N_{pix} \sum_{\nu} \nu I_{\nu}^{mod,j} \Delta \left ( ln (\nu) \right )
\end{equation}
The brightnesses $\rm I_{\nu}^{mod,j}$ in Equation \ref{eq_lj}, have been computed for each cloud, using the Dustem code and adopting the best fit parameters for the dust abundances and X$\rm _{ISRF}$ obtained in Section \ref{sec_dust_prop} from fitting the emissivity spectrum of dust associated with the molecular phase ($\rm b_{\nu}$). Due to noise in the CO data, the column density can occasionally be found negative and in those cases we did not quote the values in the tables. 

\begin{figure*}
\begin{center}
\includegraphics[width=12.5cm]{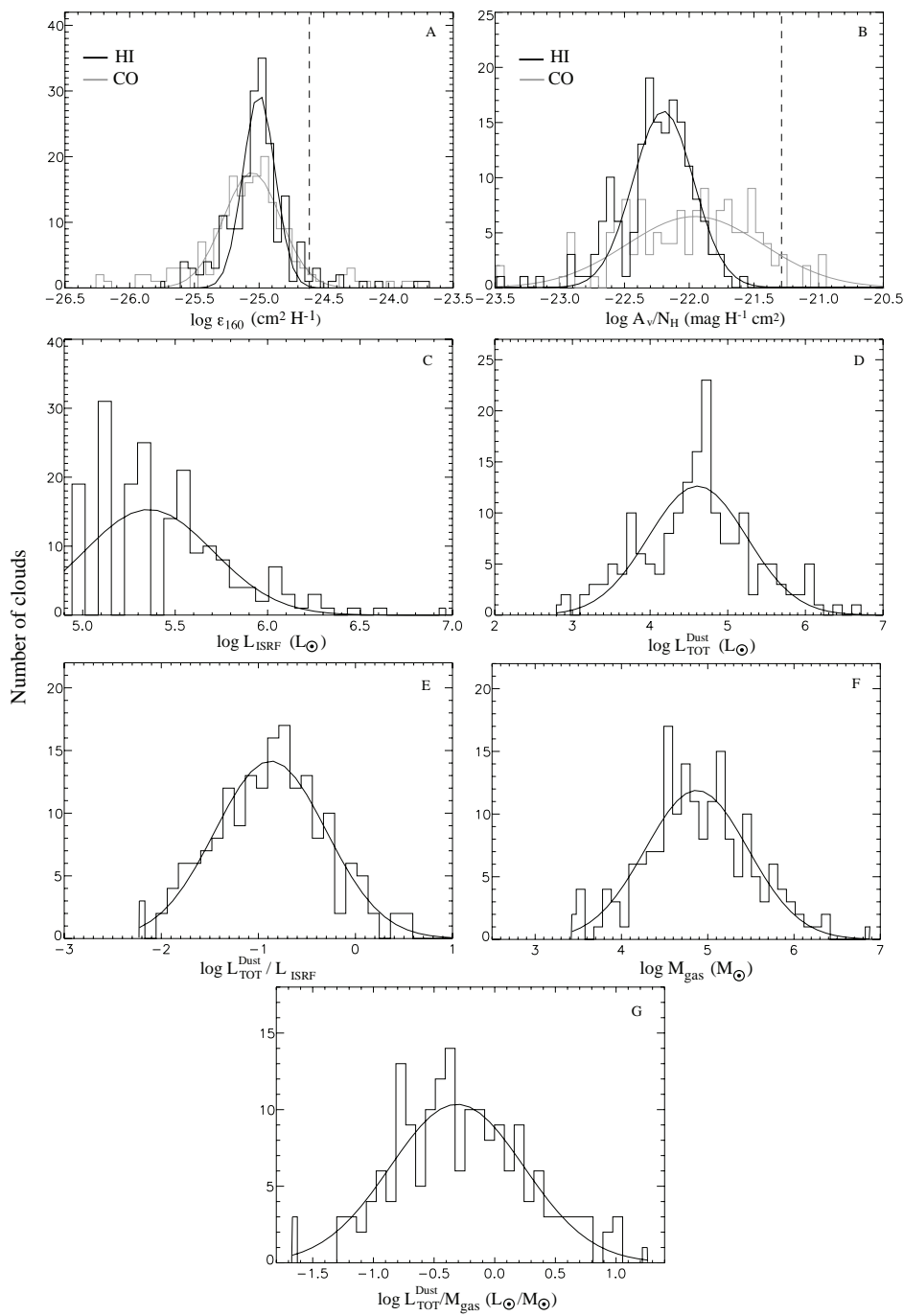}
\caption{Histograms of the logarithm of: (A) the dust emissivity at 160 $\mic$ in cm$^2$/H, in the atomic phase (black) and molecular phase (grey), (B) the extinction over column density ratio in mag H$^{-1}$ cm$^2$ in the atomic phase (black) and molecular phase (grey), (C) the incident luminosity coming from the ISRF, (D) the total luminosity emitted by dust, (E) the ratio of the total luminosity emitted by dust to the luminosity coming from the ISRF, (F) the cloud mass, and (H) the ratio of the total luminosity emitted by dust to the cloud mass. The luminosities and masses are given in unit of solar luminosity and solar mass, respectively.  The dashed vertical lines are the reference values in the solar neighborhood, quoted in Section \ref{sec_dep}.\label{fig_all_hist}}
\end{center}
\end{figure*}

\begin{figure*}
\begin{center}
\includegraphics[width=14cm]{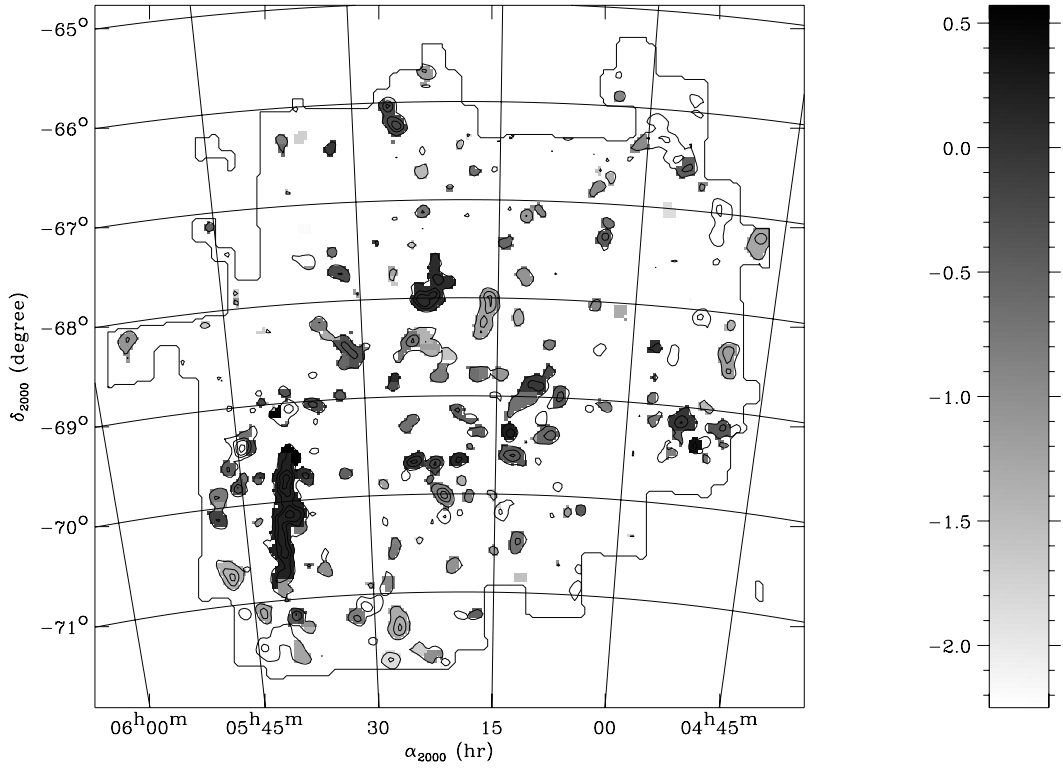}
\caption{Map of the ratio between the total luminosity emitted by dust and the ISRF luminosity ($\rm L^{Dust}_{TOT}/L_{ISRF}$), overlaid with the NANTEN $^{12}$CO (J=1-0) integrated intensity contours convolved to the 4$^{\prime}$ resolution, at 0.5, 2, 4 and 8 K km s$^{-1}$. The map is in logarithmic scale. Clouds with no values are those for which the $\rm L^{Dust}_{TOT}/L_{ISRF}$ ratio could not be derived. \label{fig_lum_rap}}
\end{center}
\end{figure*}

The ISRF luminosity and total IR luminosity values are presented in Table \ref{tab_lum}, and their histograms are shown in Figure \ref{fig_all_hist}, panels C and D, respectively. The ISRF luminosity is nearly the same for the small clouds (from cloud number 238 to 270 of Table \ref{tab_lum}), essentially due to limitations in the determination of the cloud size caused by the limited resolution of the NANTEN telescope. 
In 31$\%$ of the cases, the luminosity derived from the VSG is larger than that from the BG, whereas in our Galaxy the BG emission dominates the total dust luminosity. This result is essentially due to the 70 $\mic$ excess observed in the LMC, highlighted in \citet{Bernard08, Paradis09a}. Even if this excess resides mostly in the atomic phase, some clouds also show this excess in the molecular phase. To model the 70 $\mic$ excess in the molecular clouds of the LMC, a smaller size distribution for the VSG component compared to our Galaxy is needed as already discussed in Section \ref{sec_dust_prop}. However, in this work, we considered the same VSG size distribution as in the MW and we only allowed a change in the dust abundances, which is likely to overestimate the VSG luminosity. The total IR luminosity ranges from 6.2$\times 10^2$ to 1.8$\times 10^7$ $\rm L_{\odot}$.\\
Panel E of Figure \ref{fig_all_hist} presents the histogram of the logarithmic ratio between the total dust luminosity and the incident luminosity coming from the ISRF. Over all the clouds with finite values of this ratio, only 12 clouds show $\rm L_{TOT}^{Dust}/L_{ISRF}>1$.
In almost all cases, the dust luminosity is lower than the ISRF one, which indicates that 93$\%$ of the clouds are optically thin. The spatial distribution of this ratio is shown in Figure \ref{fig_lum_rap}.
We note that the good correspondence between Figures \ref{fig_xisrf} and \ref{fig_lum_rap} is expected due to the fact that the dust luminosity is directly proportional to the ISRF intensity.
For a collection of translucent clouds heated by a common external radiation field, we would expect a correlation betwen $\rm L_{TOT}^{Dust}/L_{ISRF}$ and A$\rm _v^{CO}$ or $\rm \tau_{160}^{CO}$. Such a correlation could not be clearly evidenced in the data. This probably indicates that energetics of most LMC clouds is set by the varying local heating from nearby stars, and not by a uniform radiation field, pervading over the whole LMC disk. Variations of the internal structure of the clouds, such as clumpiness, may also add to the dispersion.

The few clouds with $\rm L_{TOT}^{Dust}/L_{ISRF}>1$, all show warmer temperatures in the molecular phase (in the range 18.8 - 26.9 K) than most LMC clouds, and also warmer temperatures than in their surrounding atomic phase. In particular, the molecular ridge south of 30 Doradus (cloud number 197) has $\rm T_d^{CO}$=23.6 K and $\rm T_d^{HI}$=21.6 K. The high values clearly indicate active star formation, in agreement with the findings of \citet{Stanimirovic00,vanloon10}.
Almost all clouds with $\rm L_{TOT}^{Dust}/L_{ISRF}>1$ show a D/G ratio larger than the typical value given in Table \ref{tab_avg}. They also show a significantly larger VSG abundance relative to BG with respect to other LMC clouds, but no sign for an increased nor decreased PAH relative abundance. However, the increased D/G and VSG abundance affects both the molecular and the atomic phase. The molecular ridge however is an exception in this respect and does not show a modified VSG relative abundance. From the above, we tentatively conclude that clouds selected in this way are affected by star formation, and in particular show signs for an increased dust abundance and for dust processing which may have lead to the erosion of large grains into VSGs. The fact that those properties are shared by the surrounding neutral gas implies that the effects of the star formation activity are not localized to the inner molecular region, but also affect the cloud surroundings. 
  
\section{Cloud mass and star formation efficiency}
\label{sec_sf}
The cloud mass is important to understand the star formation, cloud evolution and the physics of the interstellar medium. For this study we carefully consider all the pixels of the cloud, even pixels belonging to two clouds at the same time, in order to avoid underestimating the mass. For those cloud pixels we use the average value of the $\XCO$ values of the clouds. The cloud mass is derived for each pixel i of the cloud, following \citet{Yonekura05}:
\begin{equation}
M_{gas}=\mu m_{H}D^2\Omega_{pix}\sum_i N_{H_2}^{i}=\mu m_{H}D^2\Omega_{pix} X_{CO}\sum_i W_{CO}^i
\end{equation} 
where $\mu$ is the mean molecular weight (or helium correction) taken equal to 2.8, $\rm m_{H}$ is the hydrogen atom mass, and $\rm N_{H_2}$ is the $\rm H_2$ column density. Panel F of Figure \ref{fig_all_hist} shows the histogram of the derived cloud masses. The cloud masses range from 2.7$\times 10^{3}$ to 7.5$\times 10^{6}$ M$_{\odot}$, with a most likely value deduced from the Gaussian fit equal to (7.5$^{+21.9}_{-5.6}$)$\times 10^{4}$ M$_{\odot}$.
We have compared our mass estimates with the virial masses derived by \citet{Fukui08}, which values range from 9$\times 10^{3}$ to 9$\times 10^{6}$ M$_{\odot}$. \citet{Fukui08} used the method defined by \citet{Rosolowsky06}, taking into account the effects of beam convolution and sensitivity. Our results are lower than the virial mass estimates by a factor of 1.5. This is due to the limited resolution of the map we use that underestimates the cloud size. However we have ensured consistency between our values of gas masses and luminosities (see Section \ref{sec_lum}), using the same pixel selection over the whole study. 

\begin{figure*}
\begin{center}
\includegraphics[width=8cm]{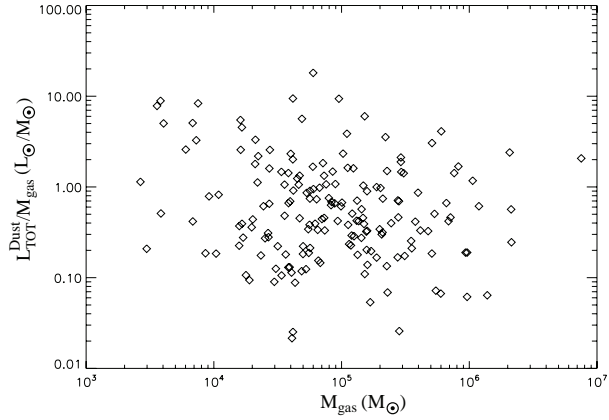}
\caption{Plot of the ratio between the total luminosity emitted by dust and the cloud mass as a function of the cloud mass. \label{fig_lir_mh2_plot}}
\end{center}
\end{figure*}

The infrared luminosity to gas mass ratio is usually used as an indicator of the SFE in a molecular cloud, with larger L/M values indicating the presence or the absence of YSOs. Figure \ref{fig_all_hist} (panel G) presents the histogram of the logarithm of the SFE. The central value deduced from the Gaussian fit corresponds to $L_{TOT}^{Dust}/M_{gas}=(0.49 ^{+1.3}_{-0.4})\,L_{\odot}/M_{\odot}$, with minimum and maximum values of 2.1$\times 10^{-2}$ and 18.1 $\rm L_{\odot}/M_{\odot}$, respectively. The maximum value corresponds to the 30 Doradus region, which is the most efficient star-forming region of the LMC. The molecular ridge, which is a huge reservoir of molecular gas, has $\rm L_{TOT}^{Dust}/M_{gas}$=2.07 $\rm L_{\odot}/M_{\odot}$, indicative of an intermediate star formation activity.
The mean ratio in the Galactic disk is 2.8 $\rm L_{\odot}/M_{\odot}$ \citep{Scoville89}. \citet{Deane94} computed this ratio in the W3 Giant Molecular Cloud located in the Perseus arm of the galaxy, using the IRAS observations. The cloud integrated average ratio is 8 $\rm L_{\odot}/M_{\odot}$, the bulk of the cloud presents lower values, in the range 0.5 - 5 $\rm L_{\odot}/M_{\odot}$, whereas the eastern ridge, with massive and active cloud cores, has the highest values ranging from 20 to 60 $\rm L_{\odot}/M_{\odot}$. They consider that clouds with a ratio close to 1.8 $\rm L_{\odot}/M_{\odot}$ are quiescent. Similarly, \citet{Mooney88} found ratios less than 1 $\rm L_{\odot}/M_{\odot}$ for quiescent clouds, and an average value around 4 $\rm L_{\odot}/M_{\odot}$ for clouds associated with HII regions.
Our median value of $\rm L_{TOT}^{Dust}/M_{gas}=0.49$ is twice lower than the typical value deduced for quiescent molecular clouds in the solar neighborhood. This is most likely due to the low metallicity of the LMC, since the IR luminosity emitted by dust is reduced as a consequence of the lower dust abundance. Therefore, we consider that $\rm L_{TOT}^{Dust}/M_{gas}<$0.5 $\rm L_{\odot}/M_{\odot}$ represent quiescent clouds in the LMC, whereas higher values should indicate star formation activity. 
In Figure \ref{fig_lir_mh2_plot} the $\rm L_{TOT}^{Dust}/M_{gas}$ ratio is plotted against the $\rm M_{gas}$. Like in the study of \citet{Sanders91} in luminous infrared galaxies, we do not see a correlation between these two quantities in the LMC. 

\begin{figure*}
\begin{center}
\includegraphics[width=8cm]{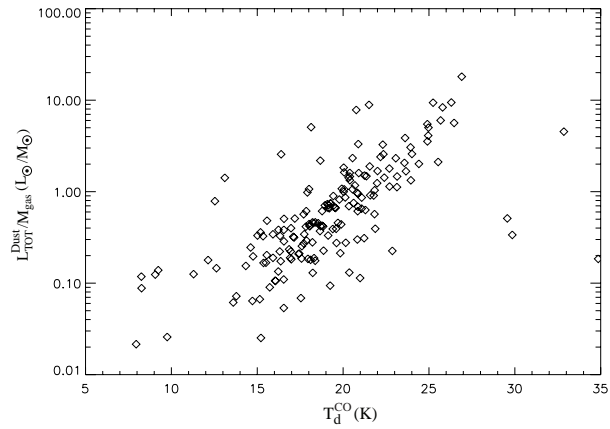}
\caption{Plot of the ratio between the total luminosity emitted by dust and the cloud mass as a function of the dust temperature in the molecular phase. \label{fig_lirmh2_tco}}
\end{center}
\end{figure*}

\begin{figure*}
\begin{center}
\includegraphics[width=14cm]{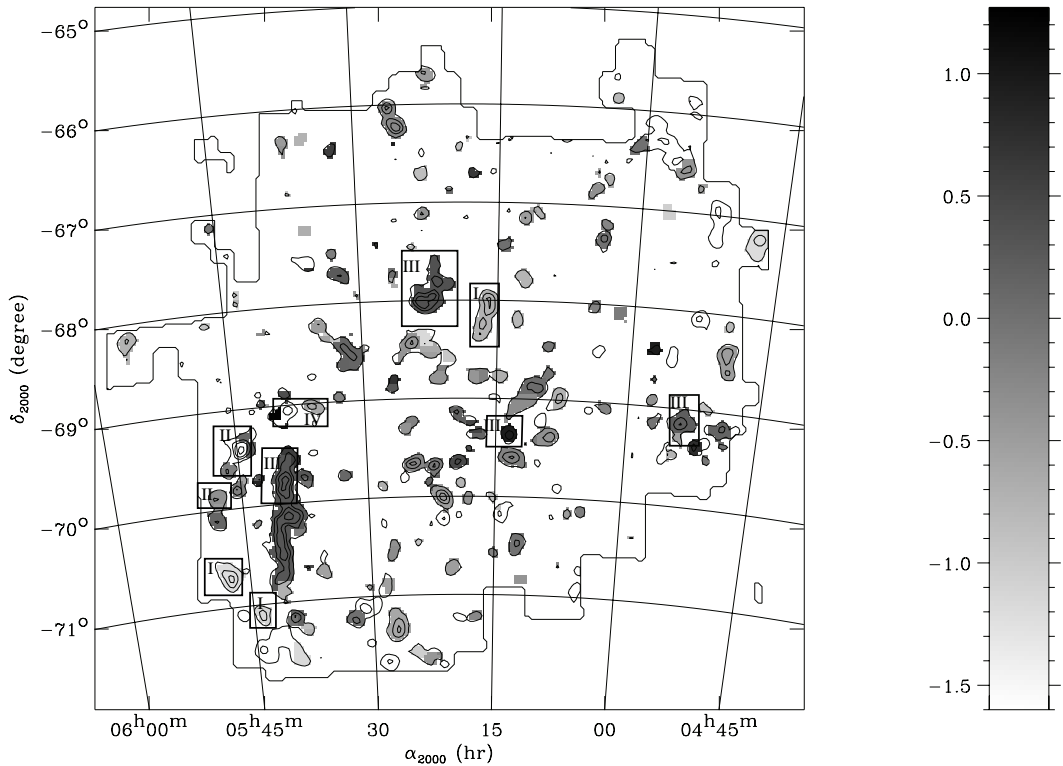}
\caption{Map of the ratio between the total luminosity emitted by dust to the cloud mass ($\rm L^{Dust}_{TOT}/M_{gas}$). The map is in logarithmic scale. The overlaid contours are described in the caption of Figure \ref{fig_lum_rap}. Clouds with no values are those for which the $\rm L^{Dust}_{TOT}/M_{gas}$ ratio could not be derived. Boxes I to IV correspond to the stages of the molecular clouds identified following the description given by \citet{Yamaguchi01}. \label{fig_lir_mh2}}
\end{center}
\end{figure*}

Figure \ref{fig_lirmh2_tco} shows $\rm L_{TOT}^{Dust}/M_{gas}$ plotted as a function of the dust temperature in the molecular phase. The observed correlation indicates that the warmer dust in the molecular phase is apparently caused by the additional heating due to star formation inside or near the cloud. Figure \ref{fig_lir_mh2} shows the distribution of the $\rm L_{TOT}^{Dust}/M_{gas}$ values across the LMC. A qualitative comparison between the spatial distribution of $\rm L_{TOT}^{Dust}/M_{gas}$ with the location of YSOs in the LMC identified by \citet{Bica96}, which spatial distribution is shown in \citet{Kawamura09} (see their Figure 3), shows a good agreement.

\begin{figure*}
\begin{center}
\includegraphics[width=8cm]{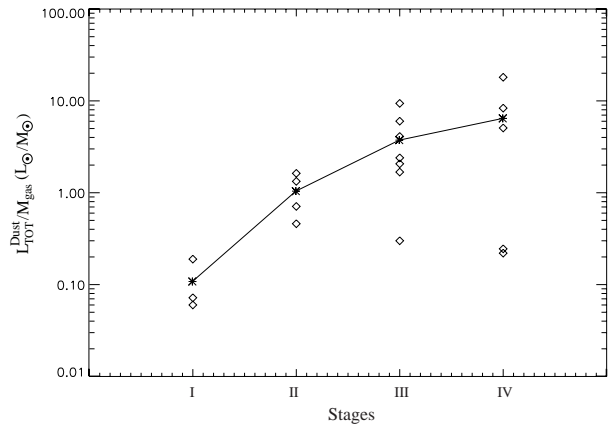}
\caption{Ratio of the total luminosity emitted by dust to the cloud mass as a function of the stage of the cloud evolution, for each selected cloud (diamonds), following the description given by \citet{Yamaguchi01}. The mean value of the ratio for each stage is shown by asterisks linked by the continuous line. \label{fig_stages}}
\end{center}
\end{figure*}

\citet{Yamaguchi01} studied the formation of stellar clusters and the evolution of 55 GMCs in the LMC, using the NANTEN telescope data. comparing the location of HII regions and young stellar clusters, they identified different stages in the evolution of GMCs (see their Figure 11). It is interesting to compare these evolution stages with the $\rm L_{TOT}^{Dust}/M_{gas}$ ratios, which is an indicator of the star formation activity. Clouds in stage I correspond to clouds with no star formation. Clouds in stage II are associated with HII regions and are the site of massive star formation. Clouds in stage III host both compact young stellar clusters and HII regions. Therefore, stages II and III represent actively star-forming clouds. Clouds in stage IV and V could correspond to the dissipation of the cloud, due to UV photons and stellar winds from the newly formed stars. In stage V, clouds are completely dissipated.  
We identified several regions with clouds in stages I to IV shown in Figure \ref{fig_lir_mh2}.
There are no clouds in stage V by definition. In Figure \ref{fig_stages}, we plot $\rm L_{TOT}^{Dust}/M_{gas}$ against the evolutionary stage for these clouds, as well as the average ratio for each stage. We can see that $\rm L_{TOT}^{Dust}/M_{gas}$ steadily increases from stage I to stage III. Although the scatter of the values increases with the evolution stage leading to a poor statistics for stage IV with 2 low values and 3 high values, the general trend indicates that
the $\rm L_{TOT}^{Dust}/M_{gas}$ ratio is a good indicator of the cloud evolution. Within this picture, LMC clouds with $\rm L_{TOT}^{Dust}/M_{gas}>0.5$ are likely to be experiencing massive star formation.

The most likely values of the luminosities and gas masses described in Sections \ref{sec_lum} and \ref{sec_sf} are summarized in Table \ref{tab_avg}.

\section{Conclusions}
\label{sec_cl}
Performing correlations between the infrared Spitzer, the IRAS and extinction data and, molecular and atomic gas tracers, we derived dust properties in the molecular region and the surrounding more diffuse parts of clouds of the LMC. We also took into account the presence of a FIR excess component of the gas, as evidenced in previous studies. This analysis is performed in connection with evaluating the star formation efficiency in the molecular clouds from FIR dust emission. Our results can be summarized as follows:

\begin{itemize}

\item{We have evidenced a slight increase of the dust temperature in the molecular phase with respect to the surrounding diffuse medium for about half the LMC molecular clouds, as opposed to what is expected for optically thick externally heated clouds. This is taken to reflect that those clouds are internally heated by star formation.}

\item{Statistically we see no significant change of the dust properties between the gas phases. Focusing on clouds with warmer dust in the molecular phase than in the atomic phase, we evidence an increase of the relative abundance of VSGs in the molecular phase.
As opposed to the situation of some of the cold molecular clouds of our Galaxy, where evidences of grain coagulation have been observed, we do not see any statistically significant sign for dust coagulation in clouds of the LMC, even those characterized by colder dust in the molecular phase than in the atomic one. This indicates that dust coagulation, if present, is not a systematic process and is not the cause for the observed lower dust temperature in some of the molecular clouds.}

\item{Our results derived from the mid-FIR correlations, and using a dust emission model, indicate that the dust abundance in the LMC is 2.2 to 2.4 times lower on average than in the solar neighborhood. Given the lower metallicity of the LMC, this is consistent with a scenario where dust abundance is in proportion to the metallicity. Extinction data indicate a significantly smaller value, 8.5 times lower than the solar neighborhood $\rm A_v/N_H$ value. The origin of this difference is unclear but could be due to a systematic bias of the star count method in the vicinity of dense clouds.} 

\item{The cloud masses derived from the FIR dust emission indicate values ranging between 2.7$\times 10^{3}$ and 7.5$\times 10^{6}$ M$_{\odot}$, with a most likely value equal to 7.5$\times 10^{4}$ M$_{\odot}$.}

\item{We find that most of the LMC clouds are translucent to the heating photons with $\rm L^{Dust}_{TOT}/M_{gas}<1$. }

\item{We find that quiescent molecular clouds have $\rm L^{Dust}_{TOT}/M_{gas}>0.5\,L_{\odot}/M_{\odot}$ whereas higher values indicate star formation activity. The maximum of star formation efficiency is found in the 30 Doradus region, with $\rm L^{Dust}_{TOT}/M_{gas}=18.1\,L_{\odot}/M_{\odot}$. 
As mentioned in some previous studies, we find that $\rm L^{Dust}_{TOT}/M_{gas}$ is not statistically correlated with $\rm M_{gas}$.
We evidence a steady increase of $\rm L^{Dust}_{TOT}/M_{gas}$ with the cloud evolutionary stage proposed by \citet{Yamaguchi01}, which confirms that this ratio is a faithful indicator of the star formation activity.}

\end{itemize}

We note that, due to the necessity of measuring the dust temperature, our conclusions are based on data smoothed to 4$^{\prime}$ resolution. Future investigations at higher angular resolution with Herschel will certainly allow one to reach more definitive conclusions.

\acknowledgments
We are very grateful to the anonymous referee for his careful reading and for his many suggestions which helped to significantly improve the quality of the manuscript. We thank Annie Hughes for her help and availability to answer to our questions. We acknowledge the use of the Dustem software package. Work on the Spitzer SAGE-LMC data has been supported by Spitzer grant 1275598 and Meixner's efforts have had additional support from NASA NAG5-12595. The production of the extinction map of the LMC was financially supported by Yamada Science Fundation for the promotion of the natural sciences (2008-1125).
\newpage
\newpage

\newpage
\newpage

\begin{deluxetable}{p{0.4cm}p{0.4cm}p{0.6cm}p{1.35cm}p{1.35cm}p{1.35cm}p{1.35cm}p{1.35cm}p{1.38cm}p{1.38cm}p{1.38cm}p{1.35cm}p{1.38cm}p{1.38cm}}
\tablecaption{Results of the IR/HI correlation in the atomic phase for all the molecular clouds. \label{tab_hi}}
\voffset=1cm
\tablewidth{620pt}
\tabletypesize{\scriptsize}
\rotate
\bigskip
\tablehead{ Cloud & RA. & DEC. &  a$_{160}$ &  a$_{100}$  & a$_{70}$ &  a$_{60}$ &a$_{25}$&  a$_{24}$  &a$_{12}$ & a$_{8}$ &  a$_{5.8}$ & a$_{4.5}$  & a$_{3.6}$   \\
&&&&&&&($10^{-1}$)& ($10^{-2}$) & ($10^{-2}$) & ($10^{-2}$) & ($10^{-2}$)  & ($10^{-3}$) &($10^{-3}$) \\
}
\startdata
     2&  71.9& -67.2&  0.11$\pm$0.01&  0.04$\pm$0.01&  0.02$\pm$0.00&  0.01$\pm$0.00&  0.03$\pm$0.00&  0.24$\pm$0.02&  0.13$\pm$0.05&  0.32$\pm$0.05&  0.03$\pm$0.02&  0.10$\pm$0.03&  0.08$\pm$0.03\\
     3&  72.2& -69.3&  0.84$\pm$0.07&  0.55$\pm$0.06&  0.36$\pm$0.04&  0.28$\pm$0.03&  0.31$\pm$0.03&  1.98$\pm$0.19&  1.65$\pm$0.12&  2.19$\pm$0.17&  0.58$\pm$0.04&  1.17$\pm$0.11&  1.41$\pm$0.11\\
     4&  72.3& -69.2&  0.96$\pm$0.06&  0.67$\pm$0.05&  0.42$\pm$0.03&  0.39$\pm$0.04&  0.42$\pm$0.07&  2.26$\pm$0.17&  2.02$\pm$0.16&  2.50$\pm$0.15&  0.76$\pm$0.06&  1.44$\pm$0.14&  1.84$\pm$0.15\\
     5&  72.4& -68.4&  0.13$\pm$0.01&  0.07$\pm$0.01&  0.04$\pm$0.00&  0.02$\pm$0.00&  0.03$\pm$0.01&  0.16$\pm$0.03&  0.38$\pm$0.07&  0.39$\pm$0.06&  0.15$\pm$0.03&  0.10$\pm$0.04&  0.30$\pm$0.05\\
     6&  72.4& -68.1&  0.14$\pm$0.01&  0.06$\pm$0.01&  0.03$\pm$0.00&  0.02$\pm$0.00&  0.02$\pm$0.00&  0.14$\pm$0.02&  0.27$\pm$0.05&  0.41$\pm$0.04&  0.14$\pm$0.02&  0.19$\pm$0.02&  0.34$\pm$0.04\\
     9&  72.6& -69.5&  0.42$\pm$0.05&  0.21$\pm$0.04&  0.18$\pm$0.02&  0.10$\pm$0.03&  0.08$\pm$0.03&  0.86$\pm$0.11&  0.90$\pm$0.10&  1.32$\pm$0.11&  0.40$\pm$0.04&  0.56$\pm$0.05&  0.63$\pm$0.06\\
    11&  72.8& -69.0&  0.81$\pm$0.04&  0.49$\pm$0.03&  0.27$\pm$0.02&  0.19$\pm$0.01&  0.14$\pm$0.02&  1.10$\pm$0.07&  1.22$\pm$0.19&  2.58$\pm$0.14&  0.82$\pm$0.05&  0.56$\pm$0.07&  0.84$\pm$0.14\\
\enddata
\tablecomments{Column 1: Cloud number as described in \citet{Fukui08}. Column 2-3: Central position of each cloud in degrees. Column 4-14: Results of the correlations  ($\rm a_{\nu}(\lambda)/X_{HI}$) in MJy/sr for 10$^{20}$ H/cm$^2$. Results that could not be derived are indicated by ' - '.}  
\end{deluxetable}

\begin{deluxetable}{p{0.4cm}p{0.4cm}p{0.6cm}p{1.35cm}p{1.35cm}p{1.35cm}p{1.35cm}p{1.35cm}p{1.38cm}p{1.38cm}p{1.38cm}p{1.35cm}p{1.38cm}p{1.38cm}}
\tablecaption{Results of the IR/CO correlation in the molecular phase for all the molecular clouds. \label{tab_co}}
\voffset=1cm
\tablewidth{620pt}
\tabletypesize{\scriptsize}
\rotate
\bigskip
\tablehead{ Cloud & RA & DEC &  b$_{160}$ &  b$_{100}$  & b$_{70}$ &  b$_{60}$ &b$_{25}$& b$_{24}$  &b$_{12}$ & b$_{8}$ &  b$_{5.8}$ & b$_{4.5}$  & b$_{3.6}$   \\
&&&&&&&($10^{-1}$)& ($10^{-2}$) & ($10^{-2}$) & ($10^{-2}$) & ($10^{-2}$)  & ($10^{-3}$) &($10^{-3}$) \\
}
\startdata
     2&  71.9& -67.2&  0.14$\pm$0.01&  0.05$\pm$0.00&  0.02$\pm$0.00&  0.01$\pm$0.00&  0.03$\pm$0.00&  0.14$\pm$0.01&  0.40$\pm$0.03&  0.49$\pm$0.03&  0.21$\pm$0.01&  $<$0.01&  0.20$\pm$0.02\\
     3&  72.2& -69.3&  1.22$\pm$0.37&  1.01$\pm$0.35&  0.53$\pm$0.19&  0.52$\pm$0.18&  0.52$\pm$0.18&  2.72$\pm$1.05&  2.97$\pm$0.66&  3.79$\pm$0.97&  1.35$\pm$0.25&  1.86$\pm$0.59&  3.21$\pm$0.61\\
     4&  72.3& -69.2&  1.63$\pm$0.12&  1.35$\pm$0.12&  0.66$\pm$0.06&  0.92$\pm$0.10&  1.59$\pm$0.17&  4.28$\pm$0.39&  4.79$\pm$0.37&  5.73$\pm$0.34&  2.30$\pm$0.14&  3.54$\pm$0.32&  4.41$\pm$0.34\\
     5&  72.4& -68.4&  0.54$\pm$0.04&  0.24$\pm$0.02&  0.09$\pm$0.01&  0.10$\pm$0.01&  0.18$\pm$0.02&  1.35$\pm$0.07&  1.47$\pm$0.20&  1.86$\pm$0.17&  0.68$\pm$0.06&  0.58$\pm$0.09&  0.58$\pm$0.14\\
     6&  72.4& -68.1&  0.32$\pm$0.06&  0.12$\pm$0.03&  0.06$\pm$0.01&  0.03$\pm$0.01&  0.11$\pm$0.01&  0.46$\pm$0.08&  0.31$\pm$0.22&  0.59$\pm$0.20&  0.14$\pm$0.09&  0.16$\pm$0.09&  0.21$\pm$0.16\\
     9&  72.6& -69.5&  0.85$\pm$0.37&  0.81$\pm$0.34&  0.41$\pm$0.16&  0.70$\pm$0.21&  1.07$\pm$0.25&  2.35$\pm$0.84&  2.36$\pm$0.77&  3.33$\pm$0.87&  1.30$\pm$0.29&  1.50$\pm$0.43&  3.14$\pm$0.52\\
    11&  72.8& -69.0&  0.17$\pm$0.11&  0.02$\pm$0.09&  $<$0.12&  $<$0.05&  0.20$\pm$0.05&  0.60$\pm$0.23&  3.92$\pm$0.61&  2.30$\pm$0.44&  0.69$\pm$0.17&  0.79$\pm$0.21&  1.98$\pm$0.43\\
\enddata
\tablecomments{Column 1: Cloud number as described in \citet{Fukui08}. Column 2-3: Central position of each cloud in degrees. Column 4-14: Results of the correlations ($\rm b_{\nu}(\lambda)/2X_{CO}$) in MJy/sr for 10$^{20}$ H/cm$^2$. Results that could not be derived are indicated by ' - '.}
\end{deluxetable}

\begin{deluxetable}{lccccccccccccc}
\tablecaption{Residuals of the IR correlation in the molecular phase for all the molecular clouds. \label{tab_res}}
\tablewidth{420pt}
\tabletypesize{\scriptsize}
\bigskip
\tablehead{ Cloud & RA & DEC &  c$_{160}$ &  c$_{100}$  & c$_{70}$ &  c$_{60}$ &c$_{25}$& c$_{24}$  &c$_{12}$ & c$_{8}$ &  c$_{5.8}$ & c$_{4.5}$  & c$_{3.6}$   \\
&&&&&&&($10^{-1}$)& ($10^{-2}$) & ($10^{-2}$) & ($10^{-2}$) & ($10^{-2}$)  & ($10^{-3}$) &($10^{-3}$) \\
}
\startdata
     2&  71.9& -67.2& -0.82& -0.22& -0.02& -0.06& -0.10& -1.66& -1.19& -1.92&  0.39& -0.79& -0.86\\
     3&  72.2& -69.3&  1.70&  2.45&  1.33&  1.19&  0.93&  8.43&  4.21&  0.90&  0.05&  2.22& -0.22\\
     4&  72.3& -69.2& -1.64& -0.41& -0.22& -0.52& -1.60& -2.17& -2.77& -4.86& -3.33& -8.82&-10.55\\
     5&  72.4& -68.4&  1.53&  0.75&  0.37&  0.21&  0.40&  0.22&  5.99&  5.85&  1.30&  1.08&  3.01\\
     6&  72.4& -68.1& -1.09& -0.60& -0.25& -0.09& -0.31& -1.37& -1.85& -3.41& -1.14& -1.36& -3.24\\
     9&  72.6& -69.5&  5.96&  5.06&  2.56&  2.67&  2.40& 10.88&  9.30& 15.98&  3.22& -0.24&  5.49\\
    11&  72.8& -69.0&  0.55&  0.28&  0.34&  0.21&  0.05&  0.17&  1.42&  2.39&  0.37&  0.13& -2.77\\
    14&  73.0& -69.4& -1.96& -3.77& -1.08&  7.01& 29.07&-10.34& 22.11& -3.09& -1.33& -3.36& -0.45\\
\enddata
\tablecomments{Column 1: Cloud number as described in \citet{Fukui08}. Column 2-3: Central position of each cloud in degrees. Column 4-14: Residuals of the correlations in MJy/sr.}
\end{deluxetable}

\begin{deluxetable}{lcccccccccc}
\tablecaption{Dust properties in the atomic phase of the clouds. \label{tab_results_hi}}
\voffset=0cm
\tablewidth{440pt}
\tabletypesize{\scriptsize}
\bigskip
\tablehead{Cloud & RA & DEC & $\rm X_{ISRF}^{HI}$ &  $\rm T_d^{HI}$& $\rm \left (\frac{Y_{VSG}}{Y_{BG}} \right )^{HI} $ & $\rm \left ( \frac{Y_{PAH}}{Y_{BG}} \right ) ^{HI}$ & $\rm (D/G)^{HI}$ & $\rm \epsilon_{160} ^{HI}$ & $\rm \left ( \frac{A_v}{N_H} \right )^{HI}$ &   $\rm A_v^0$\\
&&&&&($10^{-1}$)&$(10^{-2})$& $(10^{-3})$ &($10^{-25}$) & ($10^{-22}$) & ($10^{-2}$) \\
}
\startdata
     2&  71.9& -67.2&  0.33$\pm$0.02& 14.52$\pm$0.18 &1.52$\pm$0.34&  4.75$\pm$1.24&  1.54$\pm$0.14&  0.57$\pm$0.10&    0.42$\pm$0.16  & -6.10\\
     3&  72.2& -69.3&  1.20$\pm$0.07& 18.03$\pm$0.17  &1.98$\pm$0.31&  5.40$\pm$0.65&  3.39$\pm$0.20&  1.26$\pm$0.16&    0.27$\pm$0.12  &  0.19\\
     4&  72.3& -69.2&  1.57$\pm$0.06&  18.87$\pm$0.13 &1.74$\pm$0.23&  5.58$\pm$0.53&  3.05$\pm$0.13&  1.15$\pm$0.10&    0.25$\pm$0.11 &  -1.53\\
     5&  72.4& -68.4&  0.82$\pm$0.05&  16.92$\pm$0.16 &1.08$\pm$0.24&  8.20$\pm$1.39&  0.72$\pm$0.05&  0.26$\pm$0.04&    - &    -5.91\\
     6&  72.4& -68.1&  0.57$\pm$0.02&  15.92$\pm$0.12 &0.88$\pm$0.14&  6.63$\pm$0.92&  1.12$\pm$0.06&  0.41$\pm$0.05&   -& -0.64\\
     9&  72.6& -69.5&  0.67$\pm$0.06&  16.38$\pm$0.24 &1.46$\pm$0.38&  7.11$\pm$1.29&  2.90$\pm$0.27&  1.06$\pm$0.21&  0.24$\pm$0.05 &  -1.98  \\
    11&  72.8& -69.0&  1.30$\pm$0.03& 18.29$\pm$0.07 & 0.86$\pm$0.08&  6.82$\pm$0.47&  2.99$\pm$0.08&  1.13$\pm$0.07&   1.02$\pm$0.36   & 9.92 \\
    14&  73.0& -69.4&  0.08$\pm$0.06& 11.56$\pm$1.39 & 3.02$\pm$5.01&  5.16$\pm$7.80& 50.87$\pm$41.89& 18.52$\pm$20.11&  0.14$\pm$0.08  &  4.88  \\
\enddata
\tablecomments{Column 1: Cloud number as described in \citet{Fukui08}. Column 2-3: Central position of each cloud in degrees. Column 4: ISRF intensity. Column 5: Dust temperature in K. Column 6: VSG relative abundance compared to the BGs. Column 7: PAH relative abundance compared to the BGs. Column 8: Total dust abundance (D/G). Column 9-10: Dust emissivity at 160 $\mic$ in H/cm$^2$ and $\rm A_v/N_H$ ratio in mag/(H/cm$^{2}$), respectively. Column 11: Residual extinction in mag. Negative parameters or values that could not be derived are indicated by ' - '.}
\end{deluxetable}

\begin{deluxetable}{lcccccccccc}
\tablecaption{Dust properties in the molecular phase of the clouds. \label{tab_results_co}}
\tablewidth{480pt}
\tabletypesize{\scriptsize}
\bigskip
\tablehead{Cloud & RA & DEC &$\rm X_{CO}$ & $\rm X_{ISRF}^{CO}$ & $\rm T_d^{CO}$ & $\rm \left (\frac{Y_{VSG}}{Y_{BG}} \right )^{CO} $ & $\rm \left ( \frac{Y_{PAH}}{Y_{BG}} \right ) ^{CO}$ &$\rm (D/G)^{CO}$ & $\rm \epsilon_{160} ^{CO}$ & $\rm \left ( \frac{A_v}{N_H} \right )^{CO}$  \\
&&&($10^{20}$)&&&($10^{-1}$)&($10^{-2}$)& ($10^{-3}$) &($10^{-25}$) & ($10^{-22}$)  \\
}
\startdata
     2&  71.9& -67.2& 14.1  & 0.35$\pm$0.01& 14.72$\pm$0.07& 0.72$\pm$0.09&  9.62$\pm$0.75&  1.79$\pm$0.07&  0.64$\pm$0.07&    0.96$\pm$0.09\\ 
     3&  72.2& -69.3&   7.0 &2.62$\pm$0.42& 20.55$\pm$0.55 &1.62$\pm$0.84&  6.69$\pm$2.30&  2.57$\pm$0.44&  0.99$\pm$0.32& 2.18$\pm$0.69\\   
     4&  72.3& -69.2&    5.1&2.50$\pm$0.11&20.39$\pm$0.15 & 1.78$\pm$0.26&  7.71$\pm$0.75&  3.61$\pm$0.17&  1.37$\pm$0.13& 0.35$\pm$0.25\\   
     5&  72.4& -68.4&  3.2&0.40$\pm$0.02&  15.01$\pm$0.13&1.77$\pm$0.25&  9.09$\pm$1.19&  6.30$\pm$0.38&  2.22$\pm$0.40&   2.86$\pm$0.29\\ 
     6&  72.4& -68.1&  7.0 & 0.30$\pm$0.03& 14.33$\pm$0.25 &1.37$\pm$0.42&  3.88$\pm$1.67&  4.71$\pm$0.62&  1.74$\pm$0.46&   3.30$\pm$0.51\\
     9&  72.6& -69.5& 7.0 &3.72$\pm$0.79& 21.78$\pm$0.77 &2.13$\pm$1.24&  8.43$\pm$3.79&  1.39$\pm$0.31&  0.53$\pm$0.24&  -\\
    11&  72.8& -69.0& 8.6 &0.02$\pm$0.06&9.06$\pm$4.35 & 0.54$\pm$3.94& 18.00$\pm$120.98& 93.58$\pm$356.68& 34.80$\pm$123.68&    -\\
    14&  73.0& -69.4&  5.1&9.98$\pm$0.56& 25.68$\pm$0.24 &2.74$\pm$0.44&  4.79$\pm$0.63&  3.06$\pm$0.19&  1.31$\pm$0.14&   -\\
\enddata
\tablecomments{Column 1: Cloud number as described in \citet{Fukui08}. Column 2-3: Central position of each cloud in degrees. Column 4: $\rm X_{CO}$ conversion factor in H/cm$^2$/(K km s$^{-1}$). Column 5: ISRF intensity. Column 6: Dust temperature in K. Column 7: VSG relative abundance compared to the BGs. Column 8: PAH relative abundance compared to the BGs. Column 9: Total dust abundance (D/G). Column 10-11: Dust emissivity at 160 $\mic$ in H/cm$^2$ and $\rm A_v/N_H$ ratio in mag/(H/cm$^{2}$), respectively. Negative parameters or values that could not be derived are indicated by ' - '.}
\end{deluxetable}

\begin{deluxetable}{lcc}
\tablecaption{Most likely values of the dust properties, luminosities and gas masses, for all the clouds \label{tab_avg} }
\tablewidth{240pt}
\tabletypesize{\scriptsize}
\bigskip
\tablehead{Dust properties & Atomic phase & Molecular phase\\}
\startdata
Temperature (K) & 17.6$\pm$1.8  & 19.2$\pm$3.1  \\
Total dust abundance (D/G) & $(2.7^{+1.1}_{-1.6})\times10^{-3}$ & $(2.2^{+1.5}_{-0.9})\times 10^{-3}$  \\
$\rm Y_{PAH}/Y_{BG}$ & (5.4$^{+1.7}_{-1.3}$)10$^{-2}$&  (6.4$^{+3.0}_{-2.0}$)10$^{-2}$ \\
$\rm Y_{VSG}/Y_{BG}$ & (1.1$^{+0.6}_{-0.4}$)10$^{-1}$&  (1.4$^{+1.0}_{-0.6}$)10$^{-1}$ \\
$\rm \epsilon_{160}$ (cm$^2$/H) & (1.0$^{+0.3}_{-0.3}$)$\times$10$^{-25}$  & (8.9$^{+5.9}_{-3.5}$)$\times$10$^{-26}$  \\
$\rm A_v/N_H$ (mag/H cm$^{-2}$) & (6.3$^{+4.5}_{-2.6}$)$\times$ 10$^{-23}$  &  (1.1$^{+2.6}_{-0.8}$)$\times$10$^{-22}$ \\
\hline
Luminosities $\&$ masses & &\\
\hline
$\rm L_{ISRF}$ ($\rm L_{\odot}$) & \multicolumn{2}{c }{(2.3$^{+2.8}_{-1.3}$)$\times$ 10$^{5}$} \\
$\rm L_{TOT}^{Dust}$ ($\rm L_{\odot}$) & \multicolumn{2}{c }{(4.3$^{+14.5}_{-3.3}$)$\times$ 10$^{4}$} \\
$\rm L_{TOT}^{Dust} /L_{ISRF}$  &\multicolumn{2}{c }{ 0.14$^{+0.4}_{-0.1}$} \\
$\rm M_{gas}$ (M$\rm _{\odot}$) & \multicolumn{2}{c }{(7.5$^{+21.9}_{-5.6}$)$\times 10^{4}$}  \\
$\rm L_{TOT}^{Dust}/M_{gas}$ ($\rm L_{\odot}/M_{\odot}$)  & \multicolumn{2}{c }{$\rm 0.49^{+1.3}_{-0.4}$} \\
\hline
\enddata
\end{deluxetable}

\begin{deluxetable}{lccccccccccc}
\tablecaption{Molecular clouds luminosities and masses. \label{tab_lum}}
\tablewidth{400pt}
\tabletypesize{\scriptsize}
\bigskip
\tablehead{Cloud & RA & DEC & L$\rm _{ISRF}$  & L$\rm ^{Dust}_{TOT}$ & L$\rm _{BG}$ & L$\rm _{VSG}$ & L$\rm _{PAH}$ & $\rm \frac{L^{Dust}_{TOT}}{L_{ISRF}}$ &  $\rm M_{gas}$ & $\rm \frac{L^{Dust}_{TOT}}{M_{gas}}$ \\
&& & ($\rm 10^{6}\,L_{\odot}$) &  ($\rm 10^{4}\, L_{\odot}$) &  ($\rm 10^{4}\,L_{\odot}$) &($\rm 10^{4}\,L_{\odot}$) &($\rm 10^{3}\,L_{\odot}$) & ($\%$) &($\rm 10^{4}\,M_{\odot}$) & ($\rm M_{\odot}/L_{\odot}$)\\
}
\startdata
     2&  71.9& -67.2&    1.58&    8.84&    4.34&    2.46&   20.41&    5.59&  138.29&    0.06\\
     3&  72.2& -69.3&    0.20&    3.80&    1.98&    1.45&    3.71&   19.22&    3.58&    1.06\\
     4&  72.3& -69.2&    0.99&   42.96&   21.18&   17.17&   46.11&   43.47&   29.35&    1.46\\
     5&  72.4& -68.4&    1.98&   13.72&    4.99&    6.62&   21.15&    6.94&   41.40&    0.33\\
     6&  72.4& -68.1&    0.25&    1.02&    0.43&    0.50&    0.87&    4.12&    6.58&    0.15\\
     9&  72.6& -69.5&    0.25&    5.15&    2.42&    2.19&    5.38&   20.83&    5.67&    0.91\\
    11&  72.8& -69.0&    0.20&    0.65&    0.07&    0.19&    3.93&    3.31&    5.27&    0.12\\
    14&  73.0& -69.4&    0.35&   91.22&   42.61&   43.87&   47.45&  263.73&   15.17&    6.01\\
\enddata
\tablecomments{Column 1: Cloud number as described in \citet{Fukui08}. Column 2-3: Central position of each cloud in degrees. Column 4-8: Luminosities in unit of solar luminosity, respectively: incident luminosity coming from the ISRF, total luminosity emitted by dust, luminosity coming from the BGs, the VSGs, and the PAHs respectively. Column 9: Ratio of the total luminosity emitted by dust over the incident luminosity in percent. Column 10-11: Gas mass in unit of solar mass and ratio between the total luminosity emitted by dust and the gas mass in unit of solar mass/solar luminosity. Values that could not be derived are indicated by ' - '.}
\end{deluxetable}

\end{document}